\documentclass[11pt,a4paper]{article}
\usepackage{times}
\usepackage{jcappub}
\title{Jacobi Mapping Approach for a Precise Cosmological Weak Lensing Formalism}
\author[a]{Nastassia Grimm,} \author[a,b]{Jaiyul Yoo}
\affiliation[a]{Center for Theoretical Astrophysics and Cosmology, Institute for Computational Science, University of Z\"urich, Winterthurerstrasse 190, CH-8057, Z\"urich, Switzerland} \affiliation[b]{Physics Institute, University of Z\"urich, Winterthurerstrasse 190, CH-8057 Z\"urich, Switzerland}
\emailAdd{ngrimm@physik.uzh.ch} \emailAdd{\quad jyoo@physik.uzh.ch} 
\abstract{Cosmological weak lensing has been a highly successful and rapidly developing research field since the first detection of cosmic shear in 2000. However, it has recently been pointed out in Yoo et al.~that the standard weak lensing formalism yields gauge-dependent results and, hence, does not meet the level of accuracy demanded by the next generation of weak lensing surveys. Here, we show that the Jacobi mapping formalism provides a solid alternative to the standard formalism, as it accurately describes all the relativistic effects contributing to the weak lensing observables. 
We calculate gauge-invariant expressions for the distortion in the luminosity distance, the cosmic shear components and the lensing rotation to linear order including scalar, vector and tensor perturbations. In particular, the Jacobi mapping formalism proves that the rotation is fully vanishing to linear order. Furthermore, the cosmic shear components contain an additional term in tensor modes which is absent in the results obtained with the standard formalism. Our work provides further support and confirmation of the gauge-invariant lensing formalism needed in the era of precision cosmology.}

\begin{document}
\maketitle

\section{Introduction}
The potential of cosmological weak lensing, the deflection of light from distant sources by the large-scale structures of the universe, as a powerful cosmological probe was already recognized by theorists over half a century ago (see \cite{Gunn, Miralda1, Miralda2, Kaiser} for early work). However, a few more decades had to pass until our observational tools were sufficiently developed to measure these extremely subtle effects. In 2000, the measurement of a cosmic shear signal was reported by four independent groups (\cite{FirstDetection1, FirstDetection2, FirstDetection3, FirstDetection4}). These first observations immediately sparked great interest within the scientific community. Numerous improved observations followed soon, and cosmological weak lensing has established itself as one of the most successful and promising research fields in cosmology (see e.g.~\cite{BartelmannSchneider, ReviewMunshi, ReviewRefregier, ReviewKilbinger} for reviews).

With the next generation of weak lensing surveys, referred to as stage IV, this research field will reach its next important milestone: The ground-based observatory Large Synoptic Survey Telescope (LSST; \cite{LSST}), and the space-based missions Wide Field Infrared Survey Telescope (WFIRST; \cite{WFIRST}) and Euclid \cite{Euclid} will together cover a large fraction of the sky and measure the shape of roughly a billion galaxies with unprecedented precision. These future observations are expected to play a major role in understanding mysteries of the universe such as the nature of dark energy. However, with the high precision and the vast amount of data provided by these surveys, the scientific community is confronted with the challenging task of accounting for all sources of uncertainties to avoid false conclusions. A review of systematics in cosmic shear measurements and their theoretical interpretation which need to be brought under control to ensure the credibility of potential new findings can be found in~\cite{Mandelbaum}.

 In addition to these well-known issues, there are even more fundamental problems in the theoretical framework for cosmological weak lensing, as was pointed out recently in~\cite{newpaper}. The standard formalism used to describe cosmological weak lensing effects yields gauge-dependent results for the observables, i.e.~the convergence, the cosmic shear and the rotation. As different gauge-choices are physically indistinguishable, observable quantities have to be gauge-invariant. This discrepancy clearly shows that the standard cosmological weak lensing formalism fails to correctly account for all relativistic effects contributing to the lensing observables, and thus, conclusions based on this formalism might not be accurate. 

To some extent, flaws in the standard weak lensing formalism have already been known before. Cosmological studies measuring the magnification effect of weak lensing in fact do not measure the convergence $\kappa$, which characterizes the magnification in the standard formalism, but the distortion in the luminosity distance $\delta D$. The convergence itself is neither a gauge-invariant nor an observable quantity. One of several possibilities to calculate the luminosity distance in a perturbed FLRW universe is to apply its relation to the Jacobi map, which was first done by C.~Bonvin, R.~Durrer and M.~A.~Gasparini in 2006~\cite{Bonvin}. Although, since then, this method was applied in various other works (see e.g.~\cite{Bonvin2, CMBLensing, SecondOrderShear, UTLD, Clarkson, Clarkson2, Yamauchi}), a proof of its gauge-invariance has not been performed so far. In this work, we will prove that the Jacobi mapping formalism indeed yields a gauge-invariant expression for the distortion in the luminosity distance. To this end, we keep all ten degrees of freedom of the perturbed FLRW-metric and work with gauge-invariant perturbation quantities.  

Furthermore, we will show that the Jacobi mapping approach can be used to calculate gauge-invariant quantities for the cosmic shear and the lensing rotation, similarly to the calculation of gauge-invariant distortion in the luminosity distance $\delta D$ which replaces the convergence $\kappa$ the standard formalism. The idea to extend the Jacobi mapping approach to determine the cosmic shear components in addition to the luminosity distance was already applied in \cite{CMBLensing}, where linear order expressions coinciding with the results of the standard formalism were obtained. Moreover, this method was applied in \cite{SecondOrderShear} for second order calculations of the shear components. However, the calculations in these papers have been performed only in the Newtonian gauge for scalar modes. When tensor modes are included, we will see that the Jacobi mapping approach yields results for the cosmic shear and the rotation which are in disagreement with the standard formalism. 

This paper is structured as follows: In Section~\ref{Subsection:Metric}--\ref{Subsection:wavevector}, we introduce basic notations and discuss fundamental concepts such as the local orthonormal tetrads. These tetrads are the link between the global frame described by a FLRW-metric through which the light propagates and the local frames of the source and the observer. They are, as we will discuss, vitally important to correctly describe weak lensing effects and avoid gauge dependencies. In Section~\ref{Subsection:SF}, we will briefly summarize the standard weak lensing formalism and its gauge issues. Section~\ref{Section:JM} is the main part of this work, dedicated to establishing a precise cosmological weak lensing formalism based on the Jacobi map. We will describe how the physical lensing observables can be calculated to arbitrary order by introducing an accurate definition of the distortion matrix, ensuring that all steps of the Jacobi mapping formalism are rigorously justified. Furthermore, we will explicitly perform the linear-order calculations for the lensing observables including all scalar, vector and tensor modes and compare our results to those of the standard formalism. Finally, we will summarize and conclude our results in Section~\ref{Conclusion}.

\section{Preliminaries} \label{Preliminaries}
In this section, we first introduce in Section~\ref{Subsection:Metric} our convention for the perturbed FLRW metric and the gauge-invariant variables for the metric perturbations that will be used in this paper. In Section~\ref{Subsection:wavevector}, we introduce the local orthonormal tetrad basis, and describe how it is used to obtain the expression of the photon wavevector in FLRW coordinates from its observed value in the local observer frame. We also introduce the conformally transformed metric, which is used to significantly reduce the complexity of the Jacobi mapping formalism in Section~\ref{Section:JM}. Finally, in Section \ref{Subsection:SF} we briefly review the standard weak lensing formalism and its gauge-issues which were pointed out in \cite{newpaper}.  
\subsection{Perturbed FLRW metric and gauge-invariant variables} \label{Subsection:Metric}
The calculations in this paper are performed in a perturbed FLRW metric,
\begin{align}
\mathrm ds^2&=g_{\mu\nu}\mathrm{d}x^\mu\mathrm dx^\nu \nonumber \\
&=-a^2(\tau)(1+2\mathcal{A})\mathrm d\tau^2-2a^2(\tau)\mathcal{B}_\alpha\mathrm d\tau\mathrm dx^\alpha+a^2(\tau)\left(\delta_{\alpha\beta}+2\mathcal{C}_{\alpha\beta}\right)\mathrm dx^\alpha \mathrm dx^\beta\,,
\end{align}
where $\tau$ is the conformal time and $a(\tau)$ is the expansion scale factor. Note that we use $\mu,\nu,\rho,\dots$ to represent the 4-dimensional spacetime indices, and $\alpha,\beta,\gamma,\dots$ to represent the 3-dimensional spatial indices. The metric perturbations can be decomposed into scalar, vector and tensor perturbations:
\begin{align}
\mathcal{A}=\alpha,\qquad \mathcal{B}_\alpha=\beta_{,\alpha}+B_\alpha, \qquad \mathcal{C}_{\alpha\beta}=\varphi\delta_{\alpha\beta}+\gamma_{,\alpha\beta}+C_{(\alpha,\beta)}+C_{\alpha\beta}\,, \label{MetricDecomposition}
\end{align}
where the vector perturbations $B_\alpha$, $C_\alpha$ are divergenceless and the tensor perturbation $C_{\alpha\beta}$ is trace-free and divergenceless. In a perturbed FLRW universe, the 4-velocity $u^\mu$ of timelike flows $(u_\mu u^\mu=-1)$ also differs from its background value $\bar u^\mu=1/a\,(1,\,0,\,0,\,0)$:
\begin{align}
u^\mu\equiv\frac{1}{a}\left(1-\mathcal{A},\,\mathcal{U}^\alpha\right)\,,\qquad\mathcal{U}^\alpha\equiv-U^{,\alpha}+U^\alpha\,,\qquad \mathcal{U}_\alpha\equiv\delta_{\alpha\beta}\mathcal{U}^\beta\,,
\end{align} 
where the vector perturbation $U^\alpha$ is divergenceless. Under a gauge-transformation induced by the coordinate transformation $x^a\mapsto x^a+\xi^a$, where $\xi^a=(T,\,\mathcal{L}^\alpha)$ and $\mathcal{L}^\alpha\equiv L^{,\alpha}+L^\alpha$, the perturbation quantities transform as:
\begin{align}
&\alpha\mapsto\alpha-T'-\mathcal{H}T\,,\qquad\beta\mapsto\beta-T+L'\,,\qquad\varphi\mapsto\varphi-\mathcal{H}T\,,\qquad \gamma\mapsto\gamma-L\,,\nonumber\\
&B_\alpha\mapsto B_\alpha+L'_\alpha\,,\qquad C_\alpha\mapsto C_\alpha-L_\alpha\,,\qquad U\mapsto U-L'\,,\qquad\mathcal{U}_\alpha\mapsto\mathcal{U}_\alpha+L'_\alpha\,.
\end{align} 
 From this, we infer that the following combinations of the perturbation variables are gauge-invariant:
\begin{align}
&\alpha_\chi=\alpha-\frac{1}{a}\chi',\qquad \varphi_\chi=\varphi-H\chi,\qquad v_\chi=v-\frac{1}{a}\chi\,, \nonumber \\
&\Psi_\alpha=B_\alpha+C'_\alpha,\qquad v_\alpha=U_\alpha-B_\alpha,\qquad V_\alpha=-{v_\chi}_{,\alpha}+v_\alpha\,, \label{gmetric}
\end{align}
where $\chi\equiv a(\beta+\gamma')$ is the \textit{scalar shear} which transforms as $\chi\mapsto\chi-aT$, and $v\equiv U+\beta$ is the \textit{scalar velocity} which transforms as $v\mapsto v-T$. For future reference, we also define $\mathcal{G}^\alpha=\gamma^{,\alpha}+C^\alpha$, which transforms as $\mathcal{G}^\alpha\mapsto\mathcal{G}^\alpha-\mathcal{L}^\alpha$. The weak gravitational lensing quantities derived in this paper will be written fully in terms of the gauge-invariant variables given in \eqref{gmetric} and the gauge term $\mathcal{G}^\alpha$. Their gauge-transformation properties will thus be immediately evident.

\subsection{Orthonormal tetrads and perturbations of the photon wavevector} \label{Subsection:wavevector}
Photons emitted by some distant light source travel through the universe on null geodesics, defined by the geodesic equation $k^\mu {k^\nu}_{;\mu}=0$ and the null condition $k^\mu k_\mu=0$, where the semicolon denotes the covariant derivative with respect to the metric $g_{\mu\nu}$. The tangent vector $k^\mu$ is given by $k^\mu\equiv \mathrm dx^\mu(\Lambda)/\mathrm d\Lambda$, where $\Lambda$ is an affine parameter of the photon path. When these photons reach our observer position at some affine parameter $\Lambda_o$, we measure the photon wavevector $k^a(\Lambda_o)=\omega_o(1,\,-n^i)$, where $\omega_o$ is the angular frequency and $n^i$ is the observed photon direction. However, this measurement is performed in our local rest frame (or \textit{local Lorentz frame}), described by the Minkowski metric $\eta_{ab}$, and not in the global frame described by the perturbed FLRW metric $g_{\mu\nu}$. Note that, to distinguish them from the global FLRW coordinates, we use latin indices for the components in the local frame, where $a,b,c,\dots$ represent the 4-dimensional spacetime components and $i,j,k,\dots$ the 3-dimensional space components. 
 
The relation between the global frame and the local Lorentz frame of an observer with velocity $u^\mu$ is described by the orthonormal tetrads $e^\mu_a$ which transform the global into the local metric,
\begin{align}
\eta_{ab}=g_{\mu\nu}e_a^\mu e_b^\nu\,. \label{tetrads1}
\end{align}     
Additionally, we require that the timelike tetrad $e_0^\mu$ coincides with the observer 4-velocity, $e_0^\mu=u^\mu$. A vector $A^a$ (e.g.~the photon wavevector $k^a$) measured in the local observer frame can thus be transformed to FLRW coordinates as
\begin{align}
A^\mu=A^a e^\mu_a\,.
\end{align}
This transformation property is of fundamental importance to properly describe cosmological weak lensing observables. The concepts of the size and shape of an object, which are affected by gravitational lensing, are defined not in the global spacetime manifold described by the FLRW metric $g_{\mu\nu}$, but in the local Lorentz frame of the source. Hence, we need to transform the quantities of interest into the local Lorentz frame. 

The property in equation \eqref{tetrads1} does not uniquely define the tetrad basis. As described in \cite{newpaper}, we need to additionally take into account that the tetrads $e_i^\mu$ are four vectors in the FLRW frame, i.e.~transform as vectors under a coordinate transformation. Hence, using the notation of Section~\ref{Subsection:Metric}, their gauge-transformation property is given by
\begin{align}
e_i^0\mapsto e_i^0+\frac{1}{a}\delta_i^\alpha T_{,\alpha}\,,\qquad e_i^\alpha\mapsto e_i^\alpha+\frac{1}{a}\delta^\beta_i{L^\alpha}_{,\beta}+\frac{1}{a}\delta^\beta_i{L^{,\alpha}}_\beta+H\delta_i^\alpha T \label{tetrads2}\,.
\end{align}
 The properties in equations \eqref{tetrads1} and \eqref{tetrads2} are fulfilled by
\begin{align}
e_i^\mu=\frac{1}{a}\left(\delta_i^\beta\left(\mathcal{U}_\beta-\mathcal{B}_\beta\right),\,\delta_i^\alpha-\delta_i^\beta{p^\alpha}_\beta\right)\,,\qquad {p^\alpha}_\beta\equiv\varphi\delta^\alpha_\beta+{\mathcal{G}^\alpha}_{,\beta}+{C}^\alpha_\beta+{\epsilon^\alpha}_{\beta j}\Omega^j\,, \label{tetrads}
\end{align}
where $\Omega^j$ denotes the spatial orientation of the local frame. Note that because of the gauge-transformation property stated in equation \eqref{tetrads2}, the antisymmetric component $p_{[\alpha\beta]}=C_{[\alpha,\beta]}-{\epsilon^\alpha}_{\beta j}\Omega^j$ of the tetrads is non-vanishing. 
By applying the previous equation, we can now derive the expression for the photon wavevector $k_o^\mu$ in FLRW coordinates,
\begin{align}
k_o^\mu=\left(k^ae^\mu_a\right)_o=\frac{\omega_o}{a_o}\left(1-\mathcal{A}-n^i\delta_i^\beta\left(\mathcal{U}_\beta-\mathcal{B}_\beta\right),\,-n^i\delta_i^\alpha+\mathcal{U}^\alpha+n^i\delta_i^\beta {p^\alpha}_\beta\right)_o\,, \label{wavevectorFLRW}
\end{align}
where we require that the photon wavevector $k^a e^\mu_a$ is equal to the tangent vector $k^\mu=\mathrm dx^\mu(\Lambda)/\mathrm d\Lambda$, which uniquely fixes the parameter $\Lambda$ along the geodesic. We emphasize that the relation given in equation~\eqref{wavevectorFLRW} is valid only at the observer position since the photon direction $n^i(\Lambda)$ measured by a comoving observer at an affine parameter $\Lambda\neq\Lambda_o$ will differ from $n^i$ by a first-order quantity. 

To simplify further calculations, it is useful to express $k^\mu$ in the conformally transformed metric $\hat g_{\mu\nu}$ defined by $a^2\hat g_{\mu\nu}=g_{\mu\nu}$. As described e.g.~in~\cite{Wald}, null geodesics are invariant under conformal transformations. The photon path $x^a(\Lambda)$ is thus unaffected. However, the affine parameter $\Lambda$ is transformed into another affine parameter $\lambda$, $\mathrm d\Lambda/\mathrm d\lambda=\mathbb{C}a^2$, where the scale factor $\mathbb{C}$ is unspecified. The conformally transformed wavevector at the observer position $\hat k_o^\mu$ is given by
\begin{align}
\hat k_o^0=\left[2\mathbb{C}\pi\nu a\,\left(1-\mathcal{A}-n^\beta\left(\mathcal{U}_\beta-\mathcal{B}_\beta\right)\right)\right]_o\,,\quad\hat k_o^\alpha=\left[2\mathbb{C}\pi\nu a\,\left(-n^\alpha+\mathcal{U}^\alpha+n^\beta{p^\alpha}_\beta\right)\right]_o\,,
\end{align}
where $2\pi\nu\equiv\omega$ and $n^\alpha\equiv n^i\delta^\alpha_i$. We fix the factor $\mathbb{C}$ by requiring that at some point of the photon path specified by the affine parameter $\Lambda_p$ we have $2\mathbb{C}\pi\nu_pa_p=1$. Since the factor $2\pi\nu a$ is constant in an unperturbed FLRW universe, this choice of $\mathbb{C}$ enables us to define the perturbation variable $\widehat{\Delta\nu}$ as $2\pi\nu a\equiv 1+\Delta\nu$, where our choice of normalization is $\Delta\nu_p=0$. The perturbation variables $\delta\nu$ and $\delta n^\alpha$ for the photon wavevector can be defined as
\begin{align}
\hat k^\mu\equiv (1+\delta\nu,\,-n^\alpha-\delta n^\alpha)\,,
\end{align}
where
\begin{align}
\delta\nu_o=\left(\widehat{\Delta\nu}-\mathcal{A}-n^\alpha\left(\mathcal{U}_\alpha-\mathcal{B}_\alpha\right)\right)_o\,,\qquad\delta n_o^\alpha=\left(-n^\alpha\widehat{\Delta\nu}-\mathcal{U}^\alpha+n^\beta{p^\alpha}_\beta\right)_o\,. \label{deltanu}
\end{align}
The wavevector $\hat k^\mu$ differs from $(1,\,-n^\alpha)$ by a first-order quantity at any point of the light path, i.e.~the quantities $\delta\nu$ and $\delta n^\alpha$ are well-defined. Equation \eqref{deltanu}, which is valid only at the observer position, serves as the boundary condition. 
\subsection{Standard weak lensing formalism} \label{Subsection:SF}
The observed source position $\bar{x}_s^\mu$ of a light source observed at redshift $z$ and angular direction $(\theta,\,\phi)$ is given by
\begin{align}
\bar{x}_s^\mu=\left(\bar\tau_z,\,\bar{r}_z\sin\theta\,\cos\phi,\,\bar{r}_z\sin\theta\,\sin\phi,\,\bar{r}_z\cos\theta\right)\,,
\end{align}
where the temporal and radial coordinates are related to the redshift by
\begin{align}
\bar\tau_o-\bar\tau_z=\bar r_z=\int_0^z\frac{\mathrm dz'}{H(z')}\,,
\end{align}
and the photon path is parametrized by the conformal time, $\lambda_z=\bar\tau_z-\bar\tau_o=-\bar{r}_z$. In our real universe, however, the geodesic on which the photons travel differs from the perfectly straight path due to gravitational lensing, which causes a distortion $\delta x^\mu_s$ in the source position:
\begin{align}
x_s^\mu=\bar x^\mu_s+\delta x^\mu_s\equiv&\big(\tau_z+\Delta\tau,\,(\bar{r}_z+\delta r)\sin(\theta+\delta\theta)\cos(\phi+\delta\phi), \nonumber \\
&\phantom{(\tau_z+\Delta\tau,\,}(\bar{r}_z+\delta r)\sin(\theta+\delta\theta)\sin(\phi+\delta\phi),\,(\bar{r}_z+\delta r)\sin(\theta+\delta\theta)\big)\,.
\end{align}
For future reference, also note that the affine parameter $\lambda_s\equiv\lambda_z+\Delta\lambda_s$ and the redshift of the source are distorted, where we define the redshift distortion $\delta z$ as
\begin{align}
a_s\equiv\frac{1+\delta z}{1+z}\,. \label{defdeltaz}
\end{align}
To linear order in perturbations, the spatial source position can be written as
\begin{align}
x^\alpha_s=(\bar{r}_z+\delta r)n^\alpha+\bar{r}_z\,\delta\theta\,\theta^\alpha+\bar{r}_z\sin\theta\,\delta\phi\,\phi^\alpha\,,
\end{align}
where $n^\alpha\equiv n^i\delta^\alpha_i$ is the photon direction measured in the local rest frame of an observer at $\Lambda_o$, and $\theta^\alpha\equiv\theta^i\delta^\alpha_i$ and $\phi^\alpha\equiv\phi^i\delta^\alpha_i$ are two directions orthonormal to it,
\begin{align}
n^\alpha=\begin{pmatrix}\sin\theta\cos\phi\\\sin\theta\sin\phi\\\cos\theta\end{pmatrix}\,,\qquad\theta^\alpha=\begin{pmatrix}\cos\theta\cos\phi\\\cos\theta\sin\phi\\-\sin\theta\end{pmatrix}\,,\qquad\phi^\alpha=\begin{pmatrix}-\sin\phi\\\cos\phi\\0\end{pmatrix}\,.
\end{align}
In this paper, the quantities $n^\alpha$, $\theta^\alpha$ and $\phi^\alpha$ always refer to the directions defined at $\Lambda_o$ unless another affine parameter is specified, i.e.~$n^\alpha\equiv n^\alpha(\Lambda_o)$, $\theta^\alpha\equiv\theta^\alpha(\Lambda_o)$ and $\phi^\alpha\equiv\phi^\alpha(\Lambda_o)$. 

The distortion of the photon path affects not only the source position, but also the observed size and shape of the image. This effect is quantified by the \textit{distortion matrix} $\mathbb{D}^I{_J}$, $I=1,2$, which is also referred to as the \textit{amplification matrix}. In most literature on gravitational lensing ${\mathbb{D}^I}_J$ is defined as the $2\times 2$-dimensional projection of the Jacobian matrix of the map $\bar x_s^\alpha\mapsto x_s^\alpha$ onto the plane orthogonal to the observed photon direction $n^\alpha$, hence
\begin{align}
{\mathbb{D}^I}_J\equiv\frac{\partial\beta^I}{\partial\theta^J}\,, \label{DefDIJwrong}
\end{align}
where $\theta^I=(\theta,\,\phi)$ and $\beta^I=(\theta+\delta\theta,\,\phi+\delta\phi)$. For the geometrical interpretation of the distortion matrix, consider two light rays coming from the same source observed at an infinitesimally small angular separation $\Delta\Phi^I=(\Delta\theta,\,\sin\theta\,\Delta\phi)$ and redshift $z$. The observed spatial separation of the source is, hence, given by $\bar\xi^I_s=\bar a_s\bar r_z\Delta\Phi^I$. The Jacobian matrix of a map relates infinitesimal separations to each other, i.e.~we can write the distortion matrix $\mathbb{D}^I{_J}$ as
\begin{align}
\begin{pmatrix}\xi^\alpha_s\bar a_s\theta_\alpha\\\xi^\alpha_s\bar a_s\phi_\alpha\end{pmatrix}=\mathbb{D}^I{_J}\bar\xi^J_s\,, \label{DIJstandard}
\end{align}
where $\xi_s^\alpha=x^\alpha_s(\theta+\Delta\theta,\,\phi+\Delta\phi)-x^\alpha_s(\theta,\,\phi)$ is the separation of the FLRW coordinates at the source position. Note that in the standard formalism, the redshift distortion $\delta z$ and the distortion $\delta r$ of the radial coordinate are ignored.

The distortion matrix can be decomposed into a trace, a traceless symmetric component and a traceless anti-symmetric component,
\begin{align}
{\mathbb{D}^I}_J\equiv\begin{pmatrix}1-\kappa&0\\0&1-\kappa\end{pmatrix}-\begin{pmatrix}\gamma_1&\gamma_2\\\gamma_2&-\gamma_1\end{pmatrix}-\begin{pmatrix}0&\omega\\-\omega&0\end{pmatrix}\,,
\end{align}
where $\kappa$ is the \textit{convergence} quantifying the magnification of the image, $\gamma_1$ and $\gamma_2$ are the \textit{shear components} quantifying a distortion in shape, and $\omega$ is the \textit{rotation}. Following the definition of the distortion matrix given in equation \eqref{DefDIJwrong}, these quantities can be inferred from the expressions for $\delta\theta$ and $\delta\phi$. The true source position $x^\alpha_s$ and hence the angular distortions $\delta\theta$ and $\delta\phi$ can be obtained by integrating the photon wavevector $\hat k^\mu=\mathrm dx^\mu(\lambda)/\mathrm d\lambda$ from the observer position to the source position at the perturbed affine parameter $\lambda_s\equiv\lambda_z+\Delta\lambda_s$.

In most weak lensing literature, these calculations are performed considering only scalar perturbations and applying a certain gauge, most commonly the Newtonian gauge. To test the standard weak lensing formalism for gauge-invariance, Yoo et~al.~have calculated the expressions for $\kappa$, $\gamma_1$, $\gamma_2$ and $\omega$ resulting from the definition \eqref{DefDIJwrong} without fixing a gauge condition and also including vector and tensor modes~\cite{newpaper}. They obtained
\begin{align}
\gamma_1=&\frac{1}{2}\left(\phi_\alpha\phi_\beta-\theta_\alpha\theta_\beta\right)\left[C^{\alpha\beta}_o-\mathcal{G}^{\alpha,\beta}_s-\int_0^{\bar{r}_z}\mathrm d\bar{r}\,\frac{\partial}{\partial x_\beta}\left(\Psi^\alpha+2C^{\alpha\gamma}n_\gamma\right)\right] \nonumber \\
&+\int_0^{\bar{r}_z}\mathrm d\bar{r}\,\left(\frac{\bar{r}_z-\bar{r}}{2\bar{r}_z\bar{r}}\right)\left(\frac{\partial^2}{\partial\theta^2}-\cot\theta\frac{\partial}{\partial\theta}-\frac{1}{\sin^2\theta}\frac{\partial^2}{\partial\phi^2}\right)\left(\alpha_\chi-\varphi_\chi-n^\beta\Psi_\beta-n^\beta n^\gamma C_{\beta\gamma}\right) \label{gamma1}
\end{align}
and
\begin{align}
\gamma_2=&\frac{1}{2}\left(\theta_\alpha\phi_\beta+\theta_\beta\phi_\alpha\right)\left[-C^{\alpha\beta}_o+\mathcal{G}^{\alpha,\beta}_s+\int_0^{\bar{r}_z}\mathrm d\bar{r}\,\frac{\partial}{\partial x_\beta}\left(\Psi^\alpha+2C^{\alpha\gamma}n_\gamma\right)\right] \nonumber \\
&+\int_0^{\bar{r}_z}\mathrm d\bar{r}\,\left(\frac{\bar{r}_z-\bar{r}}{\bar{r}_z\bar{r}}\right)\frac{\partial}{\partial\theta}\left[\frac{1}{\sin\theta}\frac{\partial}{\partial\phi}\left(\alpha_\chi-\varphi_\chi-n^\beta\Psi_\beta-n^\beta n^\gamma C_{\beta\gamma}\right)\right] \label{gamma2}
\end{align}
for the shear components. For the convergence, the standard formalism yields
\begin{align}
\kappa=&\left(\frac{3}{2}C^{\alpha\beta}n_\beta-V^\alpha\right)_on_\alpha-\frac{n_\alpha\mathcal{G}_s^\alpha}{\bar{r}_z}+\frac{\widehat\nabla_\alpha\mathcal{G}_s^\alpha}{2\bar{r}_z}+\frac{n_\alpha\left(\delta x^\alpha+\mathcal{G}^\alpha\right)_o}{\bar{r}_z} \nonumber \\
&-\int_0^{\bar{r}_z}\frac{\mathrm d\bar{r}}{\bar{r}}\,\left(n_\alpha-\frac{\widehat\nabla_\alpha}{2}\right)\left(\Psi^\alpha+2C^\alpha_\beta n^\beta\right) \nonumber \\
&+\int_0^{\bar{r}_z}\mathrm d\bar{r}\,\left(\frac{\bar{r}_z-\bar{r}}{2\bar{r}_z\bar{r}}\right)\widehat{\nabla}^2\left(\alpha_\chi-\varphi_\chi-n^\beta\Psi_\beta-n^\beta n^\gamma C_{\beta\gamma}\right)\,, \label{kappa}
\end{align} 
where $\widehat\nabla_\alpha$ and $\widehat\nabla^2$ denote the angular gradient and the angular Laplace operator (see Appendix~\ref{Subsection:SphericalCoordinates}). Finally, the result for the rotation is
\begin{align}
\omega=\Omega^n_o+\frac{1}{2}\left(\theta_\alpha\phi_\beta-\phi_\alpha\theta_\beta\right)\left(\mathcal{G}_s^{\alpha,\beta}+\int_0^{\bar{r}_z}\mathrm d\bar{r}\,\frac{\partial}{\partial x_\beta}\left(\Psi^\alpha+2C^\alpha_\gamma n^\gamma\right)\right)\,, \label{omega}
\end{align}
where $\Omega^n_o=n_i\Omega^i_o$ at the observer position. These expressions are arranged such that the gauge-invariant terms are isolated from the gauge-dependent terms, i.e.~those with $\mathcal{G}^\alpha_s$. From this, it is immediately evident that none of the expressions obtained for $\gamma_1$, $\gamma_2$, $\kappa$ and $\omega$ in the standard formalism are gauge-invariant. This is a blatant contradiction to the fact that observable quantities such as the cosmic shear can not depend on physically indistinguishable gauge choices. Therefore, the standard formalism does not capture all physical effects contributing to the cosmic shear and is unsuitable for high-precision cosmological studies. 

The gauge-dependence of the expression for $\omega$ is also problematic. Several groups have argued that, although the magnitude of this effect is under debate, the rotation contributes to the B-modes of the CMB polarization power spectrum~\cite{CMBPolarization, Beppe, Fabbian, Takahashi}. Furthermore, it was proposed that the rotation could be measured in galaxy surveys by including the polarization information of the galaxy radio emission~\cite{Polarization}. Therefore, although it has not been measured yet, the lensing rotation is a physical observable and its gauge issues need to be clarified. Yoo et al.~concluded that the rotation is in fact fully vanishing for scalar, vector and tensor modes to linear order~\cite{newpaper}. We will confirm this result in the next section.  

For the convergence $\kappa$, it has already been pointed out that it is neither an observable nor a gauge-invariant quantity. Magnification effects in weak lensing are in fact quantified by the distortion in the luminosity distance $\delta D$, which is a gauge-invariant quantity closely related to the convergence~\cite{GalaxyClustering}. 
One of several possibilities to calculate $\delta D$ is to apply its relation to the Jacobi map presented in~\cite{Schneider}. However, this method has so far not been tested for gauge-invariance. In the next section, we will show that the Jacobi mapping approach indeed yields a gauge-invariant expression for the distortion in the luminosity distance and that, similarly, it can be applied to calculate gauge-invariant expressions for the shear components and the rotation.  
\section{Jacobi Mapping Approach for Cosmological Weak Lensing} \label{Section:JM}
In this section, we prove that the Jacobi mapping approach provides a precise and gauge-invariant cosmological weak lensing formalism which accounts for all relativistic effects. In Section~\ref{Subsection:JMFormalism}--\ref{Subsection:JMFormalism2} we thoroughly investigate the fully non-linear Jacobi mapping approach, paying particular attention to the subtleties of the formalism such as the discrepancy between the comoving velocity of the source and the comoving velocity of the observer parallel-transported to the source position. In Section~\ref{Subsection:JMlinear}--\ref{Subsection:Rotation}, we present the linear order results of the Jacobi mapping approach for the cosmological weak lensing quantities including not only scalar, but also vector and tensor modes. In particular, Section~\ref{Subsection:Rotation} contains a discussion of the lensing rotation, which, somewhat surprisingly, is fully vanishing to linear order. The results presented here are written fully in terms of gauge-invariant variables, which makes this work to the best of our knowledge the first proof of gauge-invariance of the Jacobi-mapping approach. Furthermore, our results coincide with those in~\cite{newpaper}, where the cosmological weak lensing quantities were derived with another method based on solving the geodesic equation for the two nearby light rays instead of solving the geodesic deviation equation.   
\subsection{Jacobi Mapping Formalism} \label{Subsection:JMFormalism}
Consider two nearby light rays that are emitted from an infinitesimally extended source at an affine parameter $\Lambda_s$ and converge at the position of an observer at an affine parameter $\Lambda_o$, where they are observed to have a small angular separation $\Delta\Phi^I=(\Delta\theta,\,\sin\theta\Delta\phi)$. 
The vector $\xi^\mu(\Lambda)$ describes the physical separation of these rays evaluated at an affine parameter $\Lambda$. In a perfectly homogeneous universe described by the FLRW metric the photons travel on straight paths, and the separation $\bar\xi_s^\mu$ at the source position can be inferred from the observed angular separation $\Delta\Phi^I$ and the observed redshift $z$, i.e.
\begin{align}
\bar\xi^\mu_s=\bar a_s\bar{r}_z\,\left(0,\,\Delta\theta\,\theta^\alpha+\sin\theta\Delta\phi\,\phi^\alpha\right)\,,
\end{align}
where the scale factor $\bar a_s$ and the radial coordinate $\bar{r}_z$ are determined by the observed redshift $z$. For a general spacetime metric, the propagation of $\xi^\mu(\Lambda)$ along the light path is described by the \textit{geodesic deviation equation},
\begin{align}
\frac{\mathrm D^2\xi^\mu(\Lambda)}{\mathrm d\Lambda^2}={R^\mu}_{\nu\rho\sigma}k^\nu k^\rho \xi^\sigma\,, \label{geodesicdeviation}
\end{align}
where $R^\mu{_{\nu\rho\sigma}}$ is the Riemann tensor.  
To determine an expression for $\xi^\mu_s$ it is useful to introduce the \textit{Jacobi map} ${\mathcal{J}^\mu}_\nu(\Lambda)$ which relates the separation $\xi^\mu(\Lambda)$ at some affine parameter $\Lambda$ to the initial value $\dot\xi^\mu_o$,
\begin{align}
\xi^\mu(\Lambda)\equiv{\mathcal{J}^\mu}_\nu(\Lambda)\dot\xi_o^\nu\,,\qquad\dot\xi^\mu_o\equiv\left.\frac{\mathrm D}{\mathrm d\Lambda}\xi^\mu(\Lambda)\right\vert_{\Lambda_o}\,.
\end{align} 
A propagation equation for ${\mathcal{J}^\mu}_\nu$ can be obtained straight-forwardly from the geodesic deviation equation \eqref{geodesicdeviation}:
\begin{align}
\frac{\mathrm D^2{\mathcal{J}^\mu}_\nu(\Lambda)}{\mathrm d\Lambda^2}=\left({R^\mu}_{\rho\sigma\tau}k^\rho k^\sigma\right){\mathcal{J}^\tau}_\nu(\Lambda)\,. \label{JacMap44}
\end{align}
The derivative $\dot\xi^\mu_o$ is used as an initial condition since the photon geodesics meet at the observer, i.e.~$\xi^\mu_o=0$, while $\dot\xi^\mu_o$ is non-zero and related to the observed angular separation as (see e.g.~\cite{UTLD}) 
\begin{align}
\dot\xi^\mu_o=-\omega_o\,(0,\,\Delta\theta\,\theta^\alpha+\sin\theta\Delta\phi\,\phi^\alpha)\,.
\end{align}
Solving the propagation equation for ${\mathcal{J}^\mu}_\nu$ would provide us with an expression for $\xi^\mu_s$. However, the vector $\xi_s^\mu$ connects two events on the global spacetime manifold described by the FLRW metric. Following the discussion in Section~\ref{Subsection:wavevector}, we need to transform $\xi_s^\mu$ to a local Lorentz frame to quantify cosmological weak lensing effects. First, note that, with the choice of parametrization specified by the condition that the photon wavevector $k^\mu=e^\mu_a k^a$ is equal to the tangent vector $\mathrm dx^\mu(\Lambda)/\mathrm d\Lambda$, the vectors $\xi^\mu_s\equiv\xi^\mu(\Lambda_s)$ and $\dot\xi^\mu_o$ live in the 2-dimensional planes which are orthogonal to the photon wavevector and to the 4-velocity at the source and the observer, respectively (see e.g.~\cite{Schneider}). In particular, this means that $\xi_s^\mu$ connects two events which are simultaneous from the point of view of the comoving observer at the source position, i.e.~in its local Lorentz frame the time component $\xi_s^t$ vanishes,
\begin{align}
\xi_s^t=\xi^\mu(\Lambda_s){u_\mu}(\Lambda_s)=\xi^\mu(\Lambda_s)e_\mu^t(\Lambda_s)=0\,.
\end{align}
Furthermore, the spatial vector $\xi_s^i$ is orthogonal to the photon direction $n_s^i$, where $k^a_s=\omega_s\,(1,\,-n_s^i)$ in local Lorentz coordinates. Hence, the separation vector $\xi_s^a=(0,\,\xi_s^i)$ is in fact a 2-dimensional object, determined by the components 
\begin{align}
\xi^I_s\equiv\xi^\mu_se^i_\mu(\Lambda_s)\Phi^I_i(\Lambda_s)\,,\qquad I=1,2, \label{defxiI}
\end{align}
 where $\Phi^i_I(\Lambda_s)=(\theta^i_s,\,\phi^i_s)$ denotes two directions orthonormal to $n^i_s$. Analogously, the vector $\dot\xi^a_o$ in the observer's rest frame is determined by the components 
\begin{align}
\dot\xi^I_o\equiv\dot\xi^\mu_o e^i_\mu(\Lambda_o)\Phi^I_i=\dot\xi^i_o\Phi^I_i\,,\qquad I=1,2\,.
\end{align}

Following these considerations, we want to define a $2\times2$-dimensional Jacobi map ${\mathfrak{D}^I}_J(\Lambda)$ as a map between 2-dimensional vectors. For that, we first define the tetrad basis $[e_I]^\mu(\Lambda_o)$, $I=1,2$, as the orthonormal tetrads which specify the 2-dimensional hypersurface orthogonal to the photon propagation direction and the 4-velocity at the observer. It is obtained by contracting the spatial tetrads ${e_i^\mu}(\Lambda_o)$ with the directions $\theta^i$ and $\phi^i$:
\begin{align}
[e_1]^\mu(\Lambda_o)\equiv{e_i^\mu}(\Lambda_o)\theta^i\,,\qquad [e_2]^\mu(\Lambda_o)\equiv{e_i^\mu}(\Lambda_o)\phi^i\,. \label{deftetrads}
\end{align}
This defines the tetrad basis $[e_I]^\mu(\Lambda_o)$, $I=1,2$, uniquely up to the rotation $\Omega_o^j$ of the local frame which corresponds to a rotation of the orthonormal basis $(n^i,\,\theta^i,\,\phi^i)$. Having specified these basis vectors in our local frame, we can set $\Omega^j_o=0$ as a gauge condition, but we choose to keep it in general. As we discuss in Section~\ref{Subsection:Rotation}, only the relation of the rotation $\Omega^j_s$, i.e.~the rotation of the local frame at the source position, to $\Omega^j_o$ is relevant, not the value of $\Omega^j_o$ itself. The tetrad basis $[e_I]^\mu(\Lambda)$ at any other affine parameter $\Lambda$ is obtained from its values at the observer position $\Lambda_o$ by parallel transport along the photon path:
\begin{align}
\frac{\mathrm D[e_I]^\mu(\Lambda)}{\mathrm d\Lambda}=\frac{\mathrm D\tilde e_i^\mu(\Lambda)}{\mathrm d\Lambda}=0\,,\qquad I=1,2\,. \label{paralleltransport}
\end{align}
The parallel-transported tetrads $\tilde e^\mu_i(\Lambda)$, fulfilling the property $[e_I]^\mu(\Lambda)=\tilde e_i^\mu\Phi^I_i$, specify the local plane of an observer with velocity $\tilde u^\mu(\Lambda)$, i.e.~the comoving velocity $u^\mu_o$ at affine parameter $\Lambda_o$ parallel-transported to the affine parameter $\Lambda$. In general, the tetrads $\tilde e^\mu_i(\Lambda)$ differ from the tetrads $e^\mu_I(\Lambda)$ given in equation~\eqref{tetrads}, which specify the local frame of an observer with comoving velocity $u^\mu(\Lambda)$.

The $2\times2$-dimensional Jacobi map ${\mathfrak{D}^I}_J\equiv\mathcal{J}^\mu{_\nu}[e^I]_\mu[e_J]^\nu$ describes the relation between the projected initial separation $\dot\xi^I_o$ and the projected separation $\tilde\xi^I$ at $\Lambda$, i.e.:
\begin{align}
\tilde\xi^I\equiv\xi^\mu[e^I]_\mu\,,\qquad\tilde\xi^I={\mathfrak{D}^I}_J(\Lambda)\dot\xi^J_o\,, \qquad \dot\xi^I_o=\dot\xi^\mu_o[e^I]_\mu=\left.\frac{\mathrm d}{\mathrm d\Lambda}\tilde\xi^I\right\vert_{\Lambda_o}\,. \label{defDIJprov}
\end{align}
The quantity $\tilde\xi^I_s$ apparently differs from the physical separation $\xi^I_s$ at the source. To obtain $\tilde\xi^I_s$, the separation vector $\xi^\mu_s$ is contracted with the tetrads $[e^I]_\mu(\Lambda_s)$, which specify the two-dimensional plane orthogonal to the photon wavevector $k^\mu_s$ and the velocity $\tilde u^\mu_s$ (note that photon wavevector is parallel-transported along the photon path, i.e.~$\tilde k^\mu_s=k^\mu_s$). However, the vectors $e^i_\mu(\Lambda_s)\Phi^I_i(\Lambda_s)$, which are contracted with $\xi^\mu_s$ to yield the 2-dimensional physical separation $\xi^I_s$ at the source, specify the plane orthogonal to the photon wavevector $k^\mu_s$ and the velocity $u^\mu_s$. The velocities $\tilde u^\mu_s$ and $u^\mu_s$ are in generally not the same, and their respective local Lorentz frames are related to each other by a Lorentz boost. Nevertheless, the quantities $\tilde\xi^I_s$ and $\xi^I_s$ coincide, since the effect of the Lorentz boost is fully absorbed in the transformation of the observed photon direction $n^i_s$, and the quantities orthogonal to it remain unaffected~\cite{CMBLensing, Ermis}, which is thoroughly proved in Appendix~\ref{Appendix:Lorentz}. Therefore, we can rewrite equation~\eqref{defDIJprov} as:
\begin{align}
\xi^I={\mathfrak{D}^I}_J(\Lambda)\dot\xi^J_o\,, \qquad \dot\xi^I_o=\dot\xi^\mu_o[e^I]_\mu=\left.\frac{\mathrm d}{\mathrm d\Lambda}\xi^I\right\vert_{\Lambda_o}\,. 
\end{align}

Using the parallel-transported tetrads $[e_I]^\mu(\Lambda_s)=\tilde e_i^\mu(\Lambda_s)\Phi^i_I$ to define the Jacobi map has two advantages over using the projected tetrads $e_i^\mu(\Lambda_s)\Phi^i_I(\Lambda_s)$ appearing in the definition of $\xi^I_s$ given in equation~\eqref{defxiI}. 
First of all, the parallel transport property \eqref{paralleltransport} enables us to rewrite the geodesic deviation equation \eqref{geodesicdeviation} into the following propagation equation for $\mathfrak{D}^I{_J}(\Lambda)$:
\begin{align}
\frac{\mathrm d^2}{\mathrm d\Lambda^2}\mathfrak{D}^I{_J}=-\mathfrak{R}^I_K\mathfrak{D}^K{_J}\,,\qquad\mbox{where}\qquad\mathfrak{R}^I_J\equiv \left({R^\mu}_{\nu\rho\sigma}k^\nu k^\sigma\right)[e^I]_\mu[e_J]^\rho\,. \label{JacMap}
\end{align}
Secondly, the tetrad basis at the observer is only defined up to the spatial rotations $\Omega^i_o$ as we have mentioned above. Analogously, the projected tetrads $e^i_\mu(\Lambda_s)\Phi^i_I(\Lambda_s)$ are only defined up to the spatial rotations $\Omega^i_s$.  
Parallel transport, however, uniquely defines the tetrad basis $[e_I]^\mu(\Lambda_s)$ once the basis $[e_I]^\mu(\Lambda_o)$ at the observer position is specified, i.e.~it fixes the relation between $\Omega^i_s$ and $\Omega^i_o$. Therefore, parallel transport is vitally important for an unambiguous and physically meaningful definition of the lensing rotation, as we will further discuss in Section~\ref{Subsection:Rotation}.
The term $\mathfrak{R}^I_J$, which describes how the Jacobi map changes along the photon path, can be decomposed into a trace $\mathfrak{R}$ and a traceless symmetric component $(\mathfrak{E}_1,\,\mathfrak{E}_2)$:
\begin{align}
\mathfrak{R}^I_J=\begin{pmatrix}\mathfrak{R}/2+\mathfrak{E}_1&\mathfrak{E}_2\\ \mathfrak{E}_2&\mathfrak{R}/2-\mathfrak{E}_1\end{pmatrix}\,.
\end{align}
Note that $\mathfrak{R}^I_J$ has no anti-symmetric component, which follows immediately from its definition given in \eqref{JacMap} and the fact that $R_{\mu\nu\rho\sigma}=R_{\rho\sigma\mu\nu}$. Furthermore, the Riemann tensor $R^\mu{_{\nu\rho\sigma}}$ can be decomposed as
\begin{align}
R^\mu{_{\nu\rho\sigma}}=C^\mu{_{\nu\rho\sigma}}-\frac{1}{2}\left(g^\mu{_{\nu\rho\tau}}R^\tau{_\sigma}+g^\mu{_{\nu\tau\sigma}}R^\tau{_\rho}\right)-\frac{R}{6}g^\mu{_{\nu\rho\sigma}}\,,\quad g_{\mu\nu\rho\sigma}\equiv g_{\mu\rho}g_{\nu\sigma}-g_{\mu\sigma}g_{\nu\rho}\,,
\end{align}
where the \textit{Weyl tensor} $C_{\mu\nu\rho\sigma}$ is the tracefree part of the Riemann tensor, i.e.~summation over any two indices vanishes. Consequently, we can write the trace $\mathfrak{R}$ as
\begin{align}
\mathfrak{R}=R^\mu{_{\nu\rho\sigma}}k^\nu k^\sigma[e^I]_\mu[e_I]^\rho=\frac{1}{2}R_{\nu\sigma}k^\nu k^\sigma\left([e^I]_\mu[e_I]^\mu\right)=R_{\nu\sigma}k^\nu k^\sigma\,, \label{mathfrakR}
\end{align}
and the traceless symmetric component as:
\begin{align}
\mathfrak{E}^I_J\equiv\begin{pmatrix}\mathfrak{E}_1&\mathfrak{E}_2\\\mathfrak{E}_2&-\mathfrak{E}_1\end{pmatrix}=C^\mu{_{\nu\rho\sigma}}k^\nu k^\sigma[e^I]_\mu[e_J]^\rho\,.
\end{align}

\subsection{Distortion matrix and conformally transformed Jacobi map} \label{Subsection:JMFormalism2}
Solving equation \eqref{JacMap} would yield an expression for the Jacobi map $\mathfrak{D}^I{_J}$ and thus for the separation $\xi^I_s$ at the source position. Consequently, this would provide us with an expression for the distortion matrix $\check{\mathbb{D}}^I{_J}$, which we define as the relation between $\xi^I_s$ and its background quantity $\bar\xi^I_s$,
\begin{align}
\xi_s^I\equiv\check{\mathbb{D}}^I{_J}\bar\xi^J_s\,. \label{DefDIJJM}
\end{align}
This definition is closely analogous to the definition of the distortion matrix $\mathbb{D}^I{_J}$ in the standard formalism, which, as stated in equation~\eqref{DIJstandard}, also relates spatial separations to each other. However, these definitons are not equivalent. The separation $\xi^I_s$ used to define $\check{\mathbb{D}}^I{_J}$ is given by $\xi^I_s=\xi^\mu_s e^i_\mu (\Lambda_s)\Phi^I_i(\Lambda_s)$, while the distortion matrix $\mathbb{D}^I{_J}$ relates the background separation $\bar\xi^I_s=\bar a_s\bar{r}_z\,\Delta\Phi^I$ to the quantities $\xi^\alpha_s\bar a_s\Phi^I_\alpha$. From the relation 
\begin{align}
\xi^\alpha_s\bar a_s\Phi^I_\alpha=\xi^\mu_s\bar e^i_\mu(\Lambda_s)\Phi^I_i(\Lambda_s)\,,\qquad\mbox{where}\qquad\bar e^i_\mu(\Lambda_s)=\bar a_s\,(0,\,\delta^i_\alpha)\,,
\end{align}
we can see that the standard formalism neglects the perturbations of the local tetrads, i.e.~the separation vector defined in global coordinates is not transformed correctly into the local rest frame of the source. This inconsistency manifests itself in the appearance of gauge-dependent terms evaluated at the source position in the expression~\eqref{gamma1}--\eqref{omega} for the lensing observables. Indeed, we will see that in the Jacobi mapping formalism these terms do not appear.    

From the relations $\dot\xi^I_o=-\omega_o\,\Delta\Phi^I$ and $\bar\xi_s^I=\bar a_s\bar r_z\,\Delta\Phi^I$, we infer that the relation between the Jacobi map ${\mathfrak{D}^I}_J$ and the distortion matrix ${\check{\mathbb{D}}^I}{_J}$ is given by
\begin{align}
{\check{\mathbb{D}}^I}{_J}=-\frac{\omega_o}{\bar{a}_s\bar{r}_z}{\mathfrak{D}^I}_J\,.
\end{align}	 
Solving the evolution equation \eqref{JacMap} for the Jacobi map would thus immediately provide us with an expression for the distortion matrix, and therefore for the convergence, shear and rotation. 

However, the evolution equation \eqref{JacMap} is a system of second-order differential equations and thus difficult to solve. A further simplification of the Jacobi mapping formalism seems necessary, which can be achieved by expressing it in the conformally transformed metric $\hat{g}_{\mu\nu}$. The conformally transformed tetrads are defined as $[\hat e_I]^\mu\equiv a[e_I]^\mu$, and 
$\hat\xi^I$ is defined such that $\xi^\mu=\hat\xi^I[\hat e_I]^\mu$ holds true, i.e.~$\hat\xi^I\equiv\xi^I/a$. The conformally transformed Jacobi map $\hat{\mathfrak{D}}^I{_J}$ is then defined by:
\begin{align}
\hat\xi^I\equiv\hat{\mathfrak{D}}^I{_J}\dot{\hat{\xi}}^J_o\,,\qquad\mbox{where}\qquad\dot{\hat\xi}^J_o\equiv\left.\frac{\mathrm d}{\mathrm d\lambda}\hat\xi^J\right\vert_{\lambda_o}\,.
\end{align}
As shown in~\cite{UTLD}, the propagation equation for $\hat{\mathfrak{D}}^I{_J}$ is analogous to the one for $\mathfrak{D}^I{_J}$,
\begin{align}
\frac{\mathrm d^2}{\mathrm d\lambda^2}\hat{\mathfrak{D}}^I{_J}=-\hat{\mathfrak{R}}^I_K\hat{\mathfrak{D}}^K{_J}\,,\qquad\mbox{where}\qquad
\hat{\mathfrak{R}}^I_J\equiv \left({\hat R^\mu}{_{\nu\rho\sigma}}\hat k^\nu \hat k^\sigma\right)[\hat e^I]_\mu[\hat e_J]^\rho\,, \label{evolutionJacMap}
\end{align}
and the initial condition is given by $\dot{\hat\xi}^I_o=-(1+\widehat{\Delta\nu}_o)\,\Delta\Phi^I$. Hence, the separation $\xi^I_s$ at the source is related to the observed angular separation $\Delta\Phi^I$ and the Jacobi map $\hat{\mathfrak{D}}^I{_J}$ as
\begin{align}
\xi_s^I=a_s\hat\xi^I(\lambda)=-a_s(1+\widehat{\Delta\nu}_o)\hat{\mathfrak{D}}^I{_J}(\lambda)\,\Delta\Phi^J\,. \label{separation}
\end{align}
Combined with the relation $\bar\xi^I_s=-\lambda_z\bar a_s\Delta\Phi^I$ for the background separation vector, we obtain the relation
\begin{align}
\check{\mathbb{D}}^I{_J}=\frac{1}{\lambda_z}\left(1+\widehat{\Delta\nu}_o\right)\left(1+\delta z\right)\hat{\mathfrak{D}}^I{_J} \label{RelDIJ}
\end{align}
between the distortion matrix $\check{\mathbb{D}}^I{_J}$ and the conformally transformed Jacobi map $\hat{\mathfrak{D}}^I{_J}$. 
To solve the system of differential equations for the Jacobi map $\hat{\mathfrak{D}}^I{_J}$ given in \eqref{evolutionJacMap}, note that the term $\hat{\mathfrak{R}}^I_J$ vanishes in the background, since the conformally transformed  metric $\hat{g}_{\mu\nu}$ is equal to the flat Minkowski metric up to perturbations. Therefore, the $n$-th order solution $\hat{\mathfrak{D}}^I{_J}{^{(n)}}$ of the Jacobi map is subject to the differential equation
\begin{align}
\frac{\mathrm d^2}{\mathrm d\lambda^2}\hat{\mathfrak{D}}^I_J{^{(n)}}=-\left(\sum_{1\le m\le n-1}\hat{\mathfrak{R}}^I_K{^{(m)}}\hat{\mathfrak{D}}^K_J{^{(n-m)}}+\hat{\mathfrak{R}}^I_K{^{(n)}}\hat{\mathfrak{D}}^K_J{^{(0)}}\right)\,, \label{JacMapGeneral}
\end{align} 
which can be solved by integrating twice provided that we already determined the solution of the Jacobi map $\hat{\mathfrak{D}}^I{_J}$ up to the $(n-1)$-th order and the expression for $\hat{\mathfrak{R}}^I{_J}$ up to the $n$-th order.
\subsection{Linear order distortion matrix} \label{Subsection:JMlinear}
In this subsection, we calculate the Jacobi map $\hat{\mathfrak{D}}^I{_J}$ and hence the distortion matrix $\check{\mathbb{D}}^I{_J}$ up to linear order. All quantities in this subsection are calculated up to a term $\mathcal{O}(2)$, which we will omit in the subsequent equations. The general evolution equation \eqref{JacMapGeneral} now takes the simple form:
\begin{align}
\frac{\mathrm d^2}{\mathrm d\lambda^2}\hat{\mathfrak{D}}^I{_J}=-\hat{\mathfrak{R}}^I_K\hat{\mathfrak{D}}^K{_J}{^{(0)}}\,.
\end{align}
This can be solved by integrating twice and applying the background relation $\hat{{\mathfrak{D}}}^I{_J}{^{(0)}}(\lambda)=\lambda\delta^I_J$. We obtain:
\begin{align}
\hat{\mathfrak{D}}^I{_J}(\lambda_s)=\lambda_s\delta^I_J+\delta\hat{\mathfrak{D}}^I{_J}=\lambda_s\delta^I_J-\lambda_s \int_0^{\lambda_s}\mathrm d\lambda\,\left(\frac{\lambda_s-\lambda}{\lambda_s\lambda}\right)\lambda^2\hat{\mathfrak{R}}^I_J(\lambda)\,. \label{solutionDIJ}
\end{align}
By substituting the solution for the Jacobi map $\hat{\mathfrak{D}}^I{_J}$ into the equation~\eqref{RelDIJ} we obtain the following expression for the distortion matrix:
\begin{align}
\check{\mathbb{D}}^I_J=\left(1+\delta z+\widehat{\Delta\nu}_o-\frac{\Delta\lambda_s}{\bar{r}_z}\right)\left(\delta^I_J-\int_0^{\lambda_s}\mathrm d\lambda\,\left(\frac{\lambda_s-\lambda}{\lambda_s\lambda}\right)\lambda^2\hat{\mathfrak{R}}^I_J(\lambda)\right)\,. \label{DeformationMatrix}
\end{align}
From this equation and the fact that $\hat{\mathfrak{R}}^I_J$ is fully symmetric, it is immediately evident that the anti-symmetric component $\check\omega$ of the distortion matrix $\check{\mathbb{D}}^I_J$ is vanishing, which confirms the result of~\cite{newpaper}. This might seem surprising to some readers, as it was believed that the vector and tensor modes would yield a non-vanishing linear order lensing rotation~\cite{Dai}. Therefore, the vanishing rotation deserves some more discussion, given in Section~\ref{Subsection:Rotation}. 

Being symmetric, the distortion matrix can be split into its trace and a tracefree symmetric matrix:
\begin{align}
\check{\mathbb{D}}^I_J=\begin{pmatrix}1-\check\kappa&0\\0&1-\check\kappa\end{pmatrix}-\begin{pmatrix}\check\gamma_1&\check\gamma_2\\\check\gamma_2&-\check\gamma_1\end{pmatrix}\,,
\end{align}
where $\check\kappa$ is given by
\begin{align}
\check\kappa=-\frac{1}{2}\left(\check{\mathbb{D}}^1_1+\check{\mathbb{D}}^2_2\right)+1=-\delta z-\widehat{\Delta\nu}_o+\frac{\Delta\lambda_s}{\bar{r}_z}+\int_0^{\lambda_s}\mathrm d\lambda\left(\frac{\lambda_s-\lambda}{\lambda_s\lambda}\right)\lambda^2\frac{1}{2}\left(\hat{\mathfrak{R}}^1_1+\hat{\mathfrak{R}}^2_2\right)\,, \label{kappaJM}
\end{align}
and the trace-free symmetric components $(\check\gamma_1,\,\check\gamma_2)$ are given by
\begin{align}
\check\gamma_1=-\frac{1}{2}\left(\check{\mathbb{D}}^1_1-\check{\mathbb{D}}^2_2\right)=\int_0^{\lambda_s}\mathrm d\lambda\left(\frac{\lambda_s-\lambda}{\lambda_s\lambda}\right)\lambda^2\frac{1}{2}\left(\hat{\mathfrak{R}}^1_1-\hat{\mathfrak{R}}^2_2\right) \label{gamma1JM}
\end{align}
and
\begin{align}
\check\gamma_2=-\frac{1}{2}\left(\check{\mathbb{D}}^1_2+\check{\mathbb{D}}^2_1\right)=-\check{\mathbb{D}}^1_2=\int_0^{\lambda_s}\mathrm d\lambda\left(\frac{\lambda_s-\lambda}{\lambda_s\lambda}\right)\lambda^2\hat{\mathfrak{R}}^1_2\,. \label{gamma2JM}
\end{align}
The first-order contribution $\check\kappa$ to the trace, which is the counterpart to the convergence $\kappa$ in the standard formalism, is determined by the trace $\hat{\mathfrak{R}}$ of the matrix $\hat{\mathfrak{R}}^I_J$ and thus by the Ricci tensor $\hat R^\mu{_\nu}$, as seen from equation~\eqref{mathfrakR}. As described e.g.~in \cite{Bonvin, UTLD, newpaper}, $\check\kappa$ is in fact equal to the distortion $\delta D$ in the luminosity distance up to a negative sign, $\check\kappa=-\delta D$. The traceless symmetric components $(\check\gamma_1,\,\check\gamma_2)$, which replace the shear components $(\gamma_1,\,\gamma_2)$ of the standard formalism, are sourced by the Weyl tensor $\hat C^\mu{_{\nu\rho\sigma}}$.  

To obtain explicit expressions for the components of the distortion matrix in terms of metric perturbations, we first need to calculate $\hat{\mathfrak{R}}^I_J(\lambda)$ defined in equation \eqref{evolutionJacMap}. Since our calculations are to linear order and the Riemann tensor ${\hat R^\mu}{_{\nu\rho\sigma}}$ of the conformally transformed metric is vanishing in the background, we only need the background values of the tetrads and the photon wavevector. Hence, $\hat{\mathfrak{R}}^I_J(\lambda)$ is given by
\begin{align}
\hat{\mathfrak{R}}^I_J=&{\hat R^\alpha}{_{\mu\beta\nu}}\hat k^\mu\hat k^\nu\Phi^I_\alpha\Phi_J^\beta \nonumber \\
=&{\hat R^\alpha}{_{0\beta 0}}\Phi^I_\alpha\Phi_J^\beta+{\hat R^\alpha}{_{\gamma\beta\delta}}n^\gamma n^\delta\Phi^I_\alpha\Phi_J^\beta-{\hat R^\alpha}{_{\gamma\beta 0}}n^\gamma\Phi^I_\alpha\Phi_J^\beta-{\hat R^\alpha}{_{0\beta\gamma}}n^\gamma\Phi^I_\alpha\Phi_J^\beta\,. \label{RIJ}
\end{align} 
Now, we need to apply the expressions for ${\hat{R}^\mu}{_{\nu\rho\sigma}}$ to obtain the result for $\hat{\mathfrak{R}}^I_J$ in terms of metric perturbations. The components of the distortion matrix are then calculated by performing the integration in the equations \eqref{kappaJM}--\eqref{gamma2JM}. This lengthy procedure is described in detail in Appendix~\ref{Appendix:JacobiMap}. We will only state the results here.
\subsection{Cosmic shear and distortion in the luminosity distance to first order} \label{Subsec:Shear}
From the expressions for the diagonal components $\check{\mathbb{D}}^1_1$ and $\check{\mathbb{D}}^2_2$ given in equation \eqref{D11D22}, we can readily compute the shear component $\check\gamma_1=(\check{\mathbb{D}}^2_2-\check{\mathbb{D}}^1_1)/2$:
\begin{align}
\check\gamma_1=&\frac{1}{2}\left(\phi_\alpha\phi_\beta-\theta_\alpha\theta_\beta\right)\left[C^{\alpha\beta}_o+C^{\alpha\beta}_s-\int_0^{\bar{r}_z}\mathrm d\bar{r}\,\frac{\partial}{\partial x_\beta}\left(\Psi^\alpha+2C^{\alpha\gamma}n_\gamma\right)\right] \nonumber \\
&+\int_0^{\bar{r}_z}\mathrm d\bar{r}\,\left(\frac{\bar{r}_z-\bar{r}}{2\bar{r}_z\bar{r}}\right)\left(\frac{\partial^2}{\partial\theta^2}-\cot\theta\frac{\partial}{\partial\theta}-\frac{1}{\sin^2\theta}\frac{\partial^2}{\partial\phi^2}\right)\left(\alpha_\chi-\varphi_\chi-n^\beta\Psi_\beta-n^\beta n^\gamma C_{\beta\gamma}\right)\,. \label{gamma1JMsolved}
\end{align}
The expression for the off-diagonal elements $\check{\mathbb{D}}^1_2=\check{\mathbb{D}}^2_1$ of the distortion matrix is given in equation \eqref{D12result}, and it is equal to the expression for $\check\gamma_2$ up to a negative sign, hence:
\begin{align}
\check\gamma_2=&\frac{1}{2}\left(\theta_\alpha\phi_\beta+\theta_\beta\phi_\alpha\right)\left[-C^{\alpha\beta}_o-C^{\alpha\beta}_s+\int_0^{\bar{r}_z}\mathrm d\bar{r}\,\frac{\partial}{\partial x_\beta}\left(\Psi^\alpha+2C^{\alpha\gamma}n_\gamma\right)\right] \nonumber \\
&+\int_0^{\bar{r}_z}\mathrm d\bar{r}\,\left(\frac{\bar{r}_z-\bar{r}}{\bar{r}_z\bar{r}}\right)\frac{\partial}{\partial\theta}\left[\frac{1}{\sin\theta}\frac{\partial}{\partial\phi}\left(\alpha_\chi-\varphi_\chi-n^\beta\Psi_\beta-n^\beta n^\gamma C_{\beta\gamma}\right)\right]\,. \label{gamma2JMsolved}
\end{align}
It is immediately evident that the expressions for $\check\gamma_1$ and $\check\gamma_2$ are gauge-invariant, as the gauge-dependent terms appearing in the equations \eqref{gamma1} and \eqref{gamma2} for the shear component $\gamma_1$ and $\gamma_2$ in the standard formalism are now absent, and all remaining terms are written fully in terms of the gauge-invariant metric perturbations. Considering only scalar modes in the Newtonian gauge, the standard formalism and the Jacobi mapping approach yield the same expressions for the shear components. However, the standard formalism does not correctly account for the effect of primordial gravitational waves, as the expressions for $\check\gamma_1$ and $\check\gamma_2$ contain a term in $C^{\alpha\beta}_s$ which is absent in the expressions for $\gamma_1$ and $\gamma_2$. Our results are fully compatible with the expressions for the shear components in~\cite{SchmidtJeong}, where the presence of the term evaluated at the source position was first pointed out, and the recent results presented in~\cite{newpaper}. 

While solving the integral in equation~\eqref{DeformationMatrix} for $\check{\mathbb{D}}^I_J$ immediately yields gauge-invariant results for $\check\gamma_1$ and $\check\gamma_2$, obtaining a gauge-invariant expression for the distortion in the luminosity distance $\delta D=-\check\kappa$ is slightly more complicated due to the contributions of the perturbation quantities $\delta z$, $\widehat{\Delta\nu}_o$ and $\Delta\lambda_o$. In Appendix~\ref{Appendix:LumDist}, we solve the temporal part of the geodesic equation to relate the perturbation of the affine parameter $\Delta\lambda_s$ to the distortion of the radial coordinate $\delta r$, and we combine these calculations with the expression~\eqref{kappaJMresult} for $\check\kappa$ resulting from the calculations of $\check{\mathbb{D}}^1_1$ and $\check{\mathbb{D}}^2_2$ in Appendix~\ref{Appendix:JacobiMap}. We obtain:
\begin{align}
\check\kappa=&-\frac{\delta r_\chi}{\bar{r}_z}-\delta z_\chi+\left(\frac{3}{2}n^\alpha n^\beta C_{\alpha\beta}-n_\alpha V^\alpha\right)_o-\left(\varphi_\chi-\frac{1}{2}n^\alpha n^\beta C_{\alpha\beta}\right)_s \nonumber \\
&-\int_0^{\bar{r}_z}\frac{\mathrm d\bar{r}}{\bar{r}}\,\left(n_\alpha-\frac{\hat\nabla_\alpha}{2}\right)\left(\Psi^\alpha+2C^\alpha_\beta n^\beta\right) \nonumber \\
&+\int_0^{\bar{r}_z}\mathrm d\bar{r}\,\left(\frac{\bar{r}_z-\bar{r}}{2\bar{r}_z\bar{r}}\right)\widehat{\nabla}^2\left(\alpha_\chi-\varphi_\chi-n^\beta\Psi_\beta-n^\beta n^\gamma C_{\beta\gamma}\right)\,, \label{checkkappa}
\end{align}
where we have defined $\delta r_\chi\equiv \delta r+n_\alpha\mathcal{G}_s^\alpha$ and $\delta z_\chi\equiv \delta z+H_s\chi_s$. The source position $x^\alpha_s$ transforms as $x^\alpha_s\,\mapsto\, x^\alpha_s+\mathcal{L}_s^\alpha$ under a gauge-transformation. Thus, $\delta r$ transforms as $\delta r\,\mapsto\,\delta r+n_\alpha\mathcal{L}^\alpha_s$, which means that $\delta r_\chi$ is a gauge-invariant quantity. Furthermore, $\delta z_\chi$ is also gauge-invariant, since $\delta z$ transforms as $\delta z\,\mapsto \delta z+\mathcal{H}_sT_s$, which follows from the definition of $\delta z$ given in equation~\eqref{defdeltaz} and the fact that $a_s$ transforms as $a_s\,\mapsto\,a_s+a_s'T_s$. Hence, equation~\eqref{checkkappa} is written fully in terms of gauge-invariant perturbation variables. Furthermore, by comparing this expression to the expression for $\kappa$ in the standard formalism, we can rewrite it as
\begin{align}
\check\kappa=\left(\kappa+\frac{n_\alpha\mathcal{G}^\alpha}{\bar{r}_z}-\frac{\hat\nabla_\alpha\mathcal{G}^\alpha}{2\bar{r}_z}\right)-\delta z_\chi-\frac{\delta r_\chi}{\bar{r}_z}+\left(\frac{1}{2}n^\alpha n^\beta C_{\alpha\beta}-\varphi_\chi\right)_s\,,
\end{align}
which coincides with the results of \cite{GalaxyClustering} and \cite{newpaper}.
\subsection{Vanishing rotation in the Jacobi mapping formalism} \label{Subsection:Rotation}
In Section~\ref{Subsection:JMlinear}, we have seen that as a straight-forward consequence of the fact that $\hat{\mathfrak{R}}^I_J$ is symmetric, the lensing rotation $\check\omega$ is fully vanishing to linear order. In particular, the gauge-invariant vector and tensor contributions integrated along the line-of-sight appearing in equation~\eqref{omega} for the rotation $\omega$ in the standard formalism are now absent. These terms imply that the spatial separation vector $\xi^\alpha$ defined in the global FLRW frame is rotated along the photon path. However, since there is no global observer the rotation within the global frame bears no physical meaning. The vanishing lensing rotation $\check\omega=0$ in the Jacobi mapping formalism implies that the image is not rotated with respect to the parallel-transported tetrad basis, as we will discuss in more detail in the following. 

First, note that the spatial rotations $\Omega^j_o$ which we left unspecified in equation~\eqref{deftetrads} yield the following contributions to the spatial components $[e_I]^\alpha(\Lambda_o)$ of the tetrads:
\begin{align}
&[e_1]^\alpha(\Lambda_o)\quad\ni\quad \theta^\beta\phi^j\epsilon^\alpha{_{\beta j}}\Omega^\phi_o+\theta^\beta n^j\epsilon^\alpha{_{\beta j}}\Omega^n_o\,,\nonumber \\
&[e_2]^\alpha(\Lambda_o)\quad\ni\quad \phi^\beta\theta^j\epsilon^\alpha{_{\beta j}}\Omega^\theta_o+\phi^\beta n^j\epsilon^\alpha{_{\beta j}}\Omega^n_o\,.
\end{align}
While the terms in $\Omega^\phi_o$ and $\Omega^\theta_o$ have a vanishing contribution to the initial separation $\dot\xi^I_o$ since it is orthogonal to the observed photon direction, the terms in $\Omega^n_o$ contribute as
\begin{align}
\dot\xi^1_o\quad\ni\quad \theta^\beta\phi^\alpha n^j\epsilon^\alpha{_{\beta j}}\Omega^n_o=-\Omega^n_o\,,\qquad\dot\xi^2_o\quad\ni\quad -\theta^\beta\phi^\alpha n^j\epsilon^\alpha{_{\beta j}}\Omega^n_o=\Omega^n_o\,.
\end{align}
Hence, the quantity $\Omega^n_o$ corresponds to a rotational degree of freedom of $\dot\xi^I_o$ within the plane orthogonal to the observed photon direction $n^i$. Similarly, constructing the projected tetrads $e^i_\mu(\Lambda_s)\Phi^I_i(\Lambda_s)$ at the source position would yield a rotational degree of freedom of the separation vector $\xi^I_s$. Consequently, if we constructed the basis at the source position instead of parallel-transporting it from the observer position, the anti-symmetric component of the distortion matrix would depend on the choice of $\Omega^i_s$ and $\Omega^i_o$, as can be seen in Section 5.1 of~\cite{newpaper} where the term $(\Omega^n_s-\Omega^n_o)$ appears in the expression for the rotation. 

In the Jacobi mapping approach, however, the rotation is fully vanishing without any degree of freedom corresponding to $\Omega^i_o$ or $\Omega^i_s$, even though we have never explicitly specified any of these two quantities. This is a consequence of the fact that in this formalism, parallel transport uniquely defines the tetrads $[e_I]^\mu(\Lambda_s)$ at the source position once the tetrads $[e_I]^\mu(\Lambda_o)$ are fixed, i.e.~it uniquely determines the relation between $\Omega^i_o$ and $\Omega^i_s$. Indeed, the parallel transport property of the tetrads led to the propagation equation of the Jacobi map $\mathfrak{D}^I{_J}$ which includes the definition of the symmetric matrix $\mathfrak{R}^I_J$. The fact that we obtain a fully symmetric distortion matrix $\mathbb{D}^I_J$ is thus a consequence of using the parallel-transported tetrads to describe the separation $\xi^I_s$ at the source position.

To define weak lensing quantities in a meaningful way, the size and shape of the source measured by an observer at affine parameter $\Lambda_o$ with respect to some observer basis should be compared to the size and shape measured by a (fictitious) observer at the source position $\Lambda_s$ with respect to the ``same'' observer basis. The notion of equality of (basis) vectors at different points of the spacetime manifold (such as the observer position and the source position at affine parameters $\Lambda_o$ and $\Lambda_s$) is defined by parallel transport. Hence, given a value of $\Omega^i_o$, any choice of $\Omega^i_s$ which differs from the one determined by parallel transport would yield an unphysical rotation sourced by the mismatch of the tetrad bases.

We conclude that the antisymmetric component of the distortion matrix calculated in the Jacobi mapping approach is indeed the physically consistent result for the lensing rotation, i.e.~it is, to linear order, fully vanishing for scalar, vector and tensor modes, which confirms the result of~\cite{newpaper}. The advantage of the Jacobi mapping approach is that it naturally includes the parallel transport of the tetrad basis while in~\cite{newpaper} it had to be explicitly calculated to correct the initially non-zero result. 

From the non-zero contribution in the standard formalism of the vector and tensor modes integrated along the line-of-sight, it follows that the image is rotated with respect to global FLRW coordinates, but not with respect to the tetrad basis. The result of the Jacobi mapping formalism implies that the tetrad basis rotates in exactly the same way with respect to global coordinates while being parallel-transported along the light geodesic, resulting in a vanishing rotation of the image with respect to the tetrad basis. A similar argument was already stated in~\cite{Dai}, where it was claimed that the B-modes induced by the rotation of the polarization vector (which follows the parallel transport equation as the tetrad basis) in the polarization spectrum of the CMB cancel out the B-modes induced by the rotation of the image which leaves the polarization unchanged. However, we strongly emphasize that the rotation of the tetrads (equal to the rotation of the polarization vector) and the rotation of the image with respect to the global coordinates are not measurable since there is no global observer. Therefore, it is incorrect to interpret these two rotations as two distinct physical effects which cancel each other out. The rotation of the image following the geodesic deviation equation with respect to the tetrad basis propagating according to the parallel transport equation comprises only one physical effect which vanishes to linear order.  

\section{Conclusion} \label{Conclusion}
In this work, we have applied the Jacobi mapping approach to calculate the distortion in the luminosity distance $\delta D$, the cosmic shear components $\check\gamma_1$ and $\check\gamma_2$, and the rotation $\check\omega$ for all scalar, vector and tensor modes. Our results are written fully in terms of gauge-invariant perturbation quantities. With this explicit check of gauge-invariance, we have provided an important confirmation of accuracy for the Jacobi mapping approach, which was previously missing even though this method is already an established lensing formalism (see e.g.~\cite{Bonvin, Bonvin2, CMBLensing, SecondOrderShear, UTLD, Clarkson, Clarkson2, Yamauchi}). 

With the Jacobi mapping approach, we obtained results for the lensing observables $\delta D$, $\check\gamma_1$, $\check\gamma_2$ and $\check\omega$ which are not compatible with the expressions for $\kappa$, $\gamma_1$, $\gamma_2$ and $\omega$ in the standard weak lensing formalism reviewed in Section~\ref{Subsection:SF}. The discrepancy between the distortion in the luminosity distance $\delta D$, an observable and gauge-invariant quantity, and the convergence $\kappa$, which is neither observable nor gauge-invariant, is already well known (see e.g.~\cite{Bonvin, Bonvin2, newpaper}).
However, for the cosmic shear and the rotation the disagreement between the Jacobi mapping formalism and the standard formalism is more surprising as the standard formalism was so far the most widely used method to compute these lensing observables. It was pointed out only recently in~\cite{newpaper} that the results for $\gamma_1$, $\gamma_2$ and $\omega$ obtained in the standard formalism are in fact gauge-dependent. Therefore, this method is unsuitable to describe the cosmic shear and the lensing rotation as these quantities are observable and consequently have to be gauge-invariant.

The difference between the Jacobi mapping approach and the standard formalism consists not only in the presence or absence of gauge-dependent terms, but is far more drastic. In fact, we showed that the lensing rotation $\check\omega$ is completely vanishing to linear order for scalar, vector and tensor perturbations. The non-vanishing integral terms appearing in the expression for the rotation in the standard formalism describe the rotation of the image with respect to the global coordinates, which has no physical meaning as there is no global observer. The physical lensing rotation $\check\omega$, however, describes the rotation of the image with respect to the parallel-transported tetrads which represent the local frames of the observer and the source. As parallel transport is naturally included in the Jacobi mapping approach, no explicit calculations of parallel-transported quantities are needed. In particular, the linear-order result $\check\omega=0$ is obtained immediately without any further calculations. This simplicity of the Jacobi mapping formalism makes it a viable option for higher-order calculations of the distortion matrix, where the antisymmetric component is not vanishing.  

We want to emphasize that with our thorough investigation of the Jacobi mapping approach, all of its assumptions are rigorously justified. In particular, we have shown that the discrepancy between the parallel-transported velocity $\tilde u^\mu_s$ and the source velocity $u^\mu_s$ does not affect the lensing observables. We conclude that $\check\omega=0$ is not arising from inconsistencies within the Jacobi mapping formalism, but is indeed the correct result for the physical lensing rotation. 

Furthermore, our calculations show that the standard formalism cannot be used to describe the contribution of primordial gravitational waves to the cosmic shear. The results from the Jacobi mapping approach yield an additional term in $C_{\alpha\beta}$ evaluated at the source position in the expressions for the shear components $\check\gamma_1$ and $\check\gamma_2$. This additional contribution of tensor modes to the weak lensing observables should be carefully evaluated to avoid false conclusions from the data of upcoming high-precision observations. The effect of tensor modes on the shear power spectra, including the term evaluated at the source position, has already been computed in~\cite{SchmidtJeong2} and \cite{Yamauchi}. These works can be extended by computing the auto- and cross-correlation fuctions of the convergence and the shear components, including the contributions of the monopole and dipole which have been ignored so far.

The results presented in this paper perfectly complement the results of~\cite{newpaper}. The method applied there is based on a different propagation equation. Here, we apply the Jacobi mapping approach based on the geodesic deviation equation, whereas the calculations in~\cite{newpaper} are based on solving the geodesic equation for two infinitesimally light rays. The fact that these independent calculations yield coinciding results is, additionally to their gauge-invariance, a strong confirmation of the accuracy of both methods.  
We believe that if the additional relativistic effects in the lensing observables are taken into account, we can expect fascinating -- and credible -- results from upcoming high-precision lensing surveys.

\section*{Acknowledgments}
We acknowledge useful discussions with Alexandre Refregier. We acknowledge support by the Swiss National Science Foundation. J.Y.~is further supported by a Consolidator Grant of the European Research Council (ERC-2015-CoG grant 680886).

\appendix
\section{Discrepancy between the Source Velocity and the parallel-transported Observer Velocity}\label{Appendix:Lorentz}
As mentioned in Section~\ref{Subsection:JMFormalism}, the source velocity $u_s^\mu$, which defines the local Lorentz frame of the source, differs in general from the velocity $\tilde u^\mu_s$, which is the velocity $u^\mu_o$ of the observer parallel-transported to the source position. Hence, the quantity $\tilde\xi^I_s$ defined in equation~\eqref{defDIJprov}, which lives in the local Lorentz frame of an observer with velocity $\tilde u^\mu_s$, apparently differs from the physical separation $\xi^I_s$ defined in the local Lorentz frame of the source. However, we have claimed that the effect of the Lorentz boost, which relates these two local frames to each other, is fully absorbed in the transformation of the observed photon direction $n^i_s$, i.e.~the quantity $\xi_s^I=\tilde\xi^I_s$ is in fact invariant under a Lorentz boost. Here, we thoroughly prove this statement. 

First, recall that $\xi^I_s$ is defined as 
\begin{align}
\xi^I_s=\xi^\mu_se^i_\mu(\Lambda_s)\Phi^I_i(\Lambda_s)\,,
\end{align} 
where $\xi^i_s=\xi^\mu_se^i_\mu(\Lambda_s)$ is the spatial separation in the rest frame of the source, and $\Phi^I_i(\Lambda_s)=(\theta^i_s,\,\phi^i_s)$ specifies two directions orthonormal to the photon direction $n^i_s$ measured in this frame. The quantity $\tilde\xi^I_s$, however, is given by 
\begin{align}
\tilde\xi^I_s=\xi^\mu_s\tilde e^i_\mu(\Lambda_s)\tilde\Phi^I_i(\Lambda_s)\,, \label{tildexi}
\end{align}
where $\tilde e^i_\mu(\Lambda_s)$ denotes the spatial tetrads $e^i_\mu(\Lambda_o)$ at the observer position, but parallel-transported to the source position. The directions $\tilde\Phi^I_i(\Lambda_s)=(\tilde\theta^i_s,\,\tilde\phi^i_s)$ specify the 2-dimensional plane orthogonal to the photon direction $\tilde n^i_s$ measured by an observer with velocity $\tilde u^\mu_s$. As the photon wavevector $k^\mu$, the velocity $\tilde u^\mu$ and the tetrad $\tilde e^\mu_i$ are parallel-transported along the photon path, we have
\begin{align}
\frac{\mathrm d}{\mathrm d\Lambda}\left(k^\mu \tilde u_\mu\right)=0
\,,\qquad \frac{\mathrm d}{\mathrm d\Lambda}\left(k^\mu\tilde e^i_\mu\right)=0\,,\qquad\frac{\mathrm d}{\mathrm d\Lambda}\left(\tilde u^\mu\tilde e^i_\mu\right)=0\,. \label{A2}
\end{align} 
This means that the observed photon direction $\tilde n^i_s$ determined by
\begin{align}
\tilde n^i=\left(\tilde u^\mu-\frac{1}{\tilde\omega}k^\mu\right)\tilde e^i_\mu\,,\qquad\mbox{where}\qquad \tilde \omega=-k^\mu\tilde u_\mu\,, \label{A3}
\end{align}
does not change along the photon path, $\tilde n^i_s=n^i$. Consequently, the orthonormal directions $\tilde\Phi^I_i$ are also equal to the respective quantities at the observer position $\Lambda_o$, i.e.~$\tilde\Phi^I_i(\Lambda_s)=\Phi^I_i$. Thus, equation~\eqref{tildexi} for $\tilde\xi^I_s$ indeed coincides with its definition stated in equation~\eqref{defDIJprov}.
 
Note that equations~\eqref{A2} and~\eqref{A3} imply that the frequency $\tilde\omega$ with respect to the observer moving with parallel-transported velocity $\tilde u^\mu$ does not change along the photon path, $\tilde \omega_s=\omega_o$. However, the frequency $\omega_o$ at the observer position appears redshifted with respect to the frequency $\omega_s$ emitted at the source. The quantity $\tilde\omega_s$ along with the quantities $\tilde n^i_s$ and $\tilde \xi^I_s$ bear no immediate physical meaning, as the velocity $\tilde u^\mu_s$ does not coincide with the source velocity $u^\mu_s$. By applying a Lorentz boost, these quantities can be transformed to match the frequency $\omega_s$, the photon direction $n^i_s$ and the physical separation $\xi^I_s$ in the rest frame of the source.

To prove that $\xi^I_s=\tilde\xi^I_s$, which assigns a physical meaning to $\tilde\xi^I_s$, we need to determine how $\xi^i_s$ and $\Phi^I_i(\Lambda_s)$ transform under a Lorentz boost $\Lambda^a{_b}$. For that, first note that the photon wavevector $k^a_s=\omega_s(1,\,-n^i_s)$ transforms as $\tilde k^a_s=\Lambda^a{_b}k^b_s$, and hence the observed photon direction $n^i_s$ transforms as
\begin{align}
\tilde n^i_s=-\Lambda_\parallel^{-1}\left(\Lambda^i{_0}-\Lambda^i{_j}n^j_s\right)\,,\qquad\mbox{where}\qquad\Lambda_\parallel=\Lambda^0{_0}-\Lambda^0{_i}n^i_s\,.
\end{align}
As a next step, we study the transformation properties of quantities perpendicular to $n^i_s$, such as the orthonormal directions $\theta^i_s$ and $\phi^i_s$. An arbitrary vector $A^a$ which is orthogonal to the photon wavevector, $A_ak^a_s=0$, can be decomposed as
\begin{align}
A^a=\left(-A_\parallel,\,A^i_\perp+n^i_sA_\parallel\right)\,,\qquad\mbox{where}\qquad A^i_\perp=P^{ij}A^j\,,\qquad P^{ij}\equiv\delta^{ij}-n^i_sn^j_s\,.
\end{align} 
After the Lorentz boost, the perpendicular component $\tilde A^i_\perp$ is given by
\begin{align}
\tilde A^i_\perp=\tilde A^i-\tilde n_s^i\tilde A_\parallel=\Lambda^i{_a}A^a-\Lambda_\parallel^{-1}\left(\Lambda^i{_0}-\Lambda^i{_j}n_s^j\right)\Lambda^0{_a}A^a\,.
\end{align}
By writing the spatial component as $A^i=A^i_\perp+n^i_sA_\parallel=A^i_\perp-n_s^iA^0$, the expression for $\tilde A^i_\perp$ can be rewritten as
\begin{align}
\tilde A^i_\perp=\left(\Lambda_\perp\right)^i{_j} A^j_\perp\,,\qquad\mbox{where}\qquad \left(\Lambda_\perp\right)^i{_j}\equiv\Lambda^i{_j}-\Lambda_\parallel^{-1}\left(\Lambda^i{_0}-\Lambda^i{_k}n^k_s\right)\Lambda^0{_j}\,. \label{defLambdaperp}
\end{align}
Hence, the perpendicular component of the Lorentz boosted vector $\tilde A^i$ is fully determined by the perpendicular component of $A^i$.

Another property which we need to show that $\tilde\xi^I_s$ and $\xi^I_s$ are equal is:
\begin{align}
\tilde X^i_\perp\tilde Y^i_\perp=\left(\Lambda_\perp\right)^i_kX^k_\perp\left(\Lambda_\perp\right)^i_lY^l_\perp=X^i_\perp Y^i_\perp\,. \label{perpcomponents}
\end{align}
To prove this, note that the coordinate transformation of two inertial frames moving with constant velocity $v^i$ with respect to each other is given by the boost matrix $\Lambda^a{_b}$ with components
\begin{align}
\Lambda^0{_0}=\gamma\,,\qquad\Lambda^0{_i}=-\gamma\beta\check v^i\,,\qquad\Lambda^i{_j}=\delta^i_j+(\gamma-1)\check v^i\check v_j\,,
\end{align}
where $\beta\equiv v=\sqrt{v^i v_i}$, $\gamma\equiv 1/\sqrt{1-\beta^2}$ and $\check v^i\equiv v^i/v$ is the unit vector in the direction of the boost. Hence, the components of the boost matrix fulfil the properties
\begin{align}
&\Lambda^i{_0}\Lambda^i{_0}=-1+\left(\Lambda^0{_0}\right)^2=-1+\gamma^2\,,\qquad\Lambda^j{_0}\Lambda^j{_i}=\Lambda^0{_0}\Lambda^0{_i}=-\gamma\beta\check v_i\,,\nonumber \\
&\Lambda^k{_i}\Lambda^k{_j}=\delta_{ij}+\Lambda^0{_i}\Lambda^0{_j}=\delta_{ij}+\gamma^2\beta^2\check v_i\check v_j\,.
\end{align}
Using these properties, one can show that
\begin{align}
\left(\Lambda_\perp\right)^i{_k}\left(\Lambda_\perp\right)^i{_l}=\delta_{kl}
\end{align} 
by applying the definition of $\left(\Lambda_\perp\right)^i{_k}$ given in~\eqref{defLambdaperp} and calculating all terms explicitly. This proves equation~\eqref{perpcomponents}. 

Having derived these general properties, we can now determine the relation between $\tilde\xi^i_s$ and $\xi^i_s$. According to the transformation property given in equation~\eqref{defLambdaperp}, the vectors $\theta^i_s$ and $\phi^i_s$ orthonormal to $n^i_s$ transform into the vectors $\tilde\theta^i_s=\theta^i$ and $\tilde\phi^i_s=\phi^i$ orthonormal to $\tilde n_s^i=n^i$ as
\begin{align}
\tilde\theta^i_s=\left(\Lambda_\perp\right)^i{_j}\theta^j_s\,,\qquad\tilde\phi^i_s=\left(\Lambda_\perp\right)^i{_j}\phi^j_s\,.
\end{align} 
Finally, equation~\eqref{perpcomponents} yields
\begin{align}
\tilde\xi^I_s=\left(\tilde\xi^i\tilde\Phi^I_i\right)_s=\left(\tilde\xi^i_\perp\tilde\Phi^I_i\right)_s=\left(\xi^i_\perp\Phi^I_i\right)_s=\left(\xi^i\Phi^I_i\right)_s=\xi^I_s\,,
\end{align}
which completes our proof that the physical separation $\xi^I_s$ of the source is unaffected by a Lorentz boost. Therefore, the fact that the Jacobi mapping approach is based on the parallel transport of the tetrads, and hence of the velocity, does not lead to any inconsistencies.
\section{Calculations of the Distortion Matrix in the Jacobi Mapping Approach} \label{Appendix:JacobiMap}
In this section, we perform the calculations of the linear-order distortion matrix $\check{\mathbb{D}}^I_J$ which describes how an image is affected by cosmological weak lensing. We first calculate in Section~\ref{Subsection:Riemann} the expressions for the conformally transformed Riemann tensor $\hat{R}^\mu{_{\nu\sigma\rho}}$. Then, we state in Section~\ref{Subsection:SphericalCoordinates} some useful relations in spherical coordinates which will be applied extensively in the subsequent sections. In Section~\ref{Subsection:D11D22}, we calculate the diagonal components $\check{\mathbb{D}}^1_1$ an $\check{\mathbb{D}}^2_2$ of the distortion matrix by first calculating $\hat{\mathfrak{R}}^1_1$ and $\hat{\mathfrak{R}}^2_2$ from the expressions for the Riemann tensor $\hat R^\mu{_{\nu\sigma\rho}}$ and then calculating the integral in equation~\eqref{DeformationMatrix}. Finally, in Section~\ref{Subsection:D12}, we perform these calculations for the off-diagonal components $\check{\mathbb{D}}^1_2=\check{\mathbb{D}}^2_1$. 

\subsection{Riemann tensor in the conformally transformed metric} \label{Subsection:Riemann}
For the calculation of the distortion matrix $\check{\mathbb{D}}^I_J$, we need to know the components of the Riemann tensor ${\hat{R}^\mu}{_{\nu\sigma\rho}}$ in the conformally transformed metric $\hat{g}_{\mu\nu}$. For that, first recall that the Christoffel symbols $\hat\Gamma^\sigma{_{\mu\nu}}$ are related to the metric by
\begin{align}
\hat\Gamma^\sigma{_{\mu\nu}}=\frac{1}{2}\hat g^{\sigma\rho}\left(\hat g_{\nu\rho,\mu}+\hat g_{\mu\rho,\nu}-\hat g_{\mu\nu,\rho}\right)\,,
\end{align}
which yields
\begin{align}
&\hat\Gamma^0{_{00}}=\mathcal{A}'\,,\qquad\hat\Gamma^0{ _{0\alpha}}=\mathcal{A}_{,\alpha}\,,\qquad\hat\Gamma^0 {_{\alpha\beta}}=\mathcal{B}_{(\alpha,\beta)}+\mathcal{C}'_{\alpha\beta}\,,\qquad\hat\Gamma^\alpha{_{00}}=\mathcal{A}^{,\alpha}-{\mathcal{B}^\alpha}'\,, \nonumber \\
&\hat\Gamma^\alpha{_{0\beta}}=\frac{1}{2}\left({\mathcal{B}_\beta}^{,\alpha}-{\mathcal{B}^\alpha}_{,\beta}\right)+{\mathcal{C}^\alpha_\beta}'\,,\qquad\hat\Gamma^\alpha{_{\beta\gamma}}=2{\mathcal{C}^\alpha}_{(\beta,\gamma)}-{\mathcal{C}_{\beta\gamma}}^{,\alpha}\,.
\end{align}
The Riemann tensor ${\hat R^\mu}{_{\nu\rho\sigma}}$ is related to the Christoffel symbols by
\begin{align}
{\hat R^\mu{_{\nu\rho\sigma}}}=\hat\Gamma^\mu{_{\nu\sigma,\rho}}-\hat\Gamma^\mu{_{\rho\sigma,\nu}}+\hat\Gamma^\kappa{_{\nu\sigma}}\hat\Gamma^\mu{_{\kappa\rho}}-\hat\Gamma^\kappa{_{\rho\sigma}}\hat\Gamma^\mu{_{\kappa\nu}}\,.
\end{align}
From this, we obtain that the Riemann tensor components which will appear in further calculations are given by
\begin{align}
\hat{R}^\alpha{_{0\beta0}}&={\mathcal{A}}^{,\alpha}{_\beta}-\frac{1}{2}\left({{\mathcal{B}}_\beta}{^{,\alpha}}+{{\mathcal{B}}^\alpha}{_{,\beta}}\right)'-{{\mathcal{C}}^\alpha_\beta}{''}\, \nonumber \\
{\hat R^\alpha}{_{0\beta\gamma}}&=-{{\mathcal{B}}_{[\beta}}{^{,\alpha}}{_{\gamma]}}-2{{\mathcal{C}}^\alpha}{_{[\beta,\gamma]}}'\,, \nonumber \\
\hat{R}^\alpha{_{\beta\gamma0}}&=\frac{1}{2}\left({{\mathcal{B}}_\beta}{^{,\alpha}}-{{\mathcal{B}}^\alpha} {_{,\beta}}\right)_{,\gamma}-{\mathcal{C}}^\alpha_{\gamma,\beta}{'}+{{\mathcal{C}}{_{\beta\gamma}}}{^{,\alpha}}{'}\,, \nonumber \\
{\hat R^\alpha}{_{\beta\gamma\delta}}&=2{\mathcal{C}}^\alpha{_{[\delta,\gamma]\beta}}+2{\mathcal{C}}_{\beta[\gamma,\delta]}{^\alpha}\,.
\end{align}
The components $\hat R^0{_{\alpha0\beta}}$ and $\hat R^0{_{\alpha\beta\gamma}}$ are also non-vanishing, but will be of no importance for further calculations.
\subsection{Useful relations in spherical coordinates}\label{Subsection:SphericalCoordinates}
Here, we briefly review some basic properties of the spherical coordinates $(r,\,\theta,\,\phi)$ which are vitally important for the calculation of the distortion matrix. First, recall that we have defined two directions $\theta^\alpha$ and $\phi^\alpha$ orthogonal to the observed photon direction $n^\alpha$ in Section~\ref{Subsection:SF}. These three vectors provide a basis of orthonormal unit vectors in spherical coordinates. The gradient can be written in terms of these vectors as
\begin{align}
\nabla_\alpha=n_\alpha\frac{\partial}{\partial r}+\frac{1}{r}\left(\theta_\alpha\frac{\partial}{\partial\theta}+\frac{1}{\sin\theta}\phi_\alpha\frac{\partial}{\partial\phi}\right)\equiv n_\alpha\frac{\partial}{\partial r}+\frac{1}{r}\widehat\nabla_\alpha\,,
\end{align} 
and the Laplacian operator can be written as
\begin{align}
\Delta=&\frac{\partial^2}{\partial r^2}+\frac{2}{r}\frac{\partial}{\partial r}+\frac{1}{r^2}\left[\left(\cot\theta+\frac{\partial}{\partial\theta}\right)\frac{\partial}{\partial\theta}+\frac{1}{\sin^2\theta}\frac{\partial^2}{\partial\phi^2}\right] \nonumber \\
\equiv&\frac{\partial^2}{\partial r^2}+\frac{2}{r}\frac{\partial}{\partial r}+\frac{1}{r^2}\widehat\nabla^2\,.
\end{align}
From the expression of the gradient, it follows that given any scalar function $Y$, we have
\begin{align}
n_\alpha Y^{,\alpha}=\frac{\partial}{\partial r}Y\,,\qquad\theta_\alpha Y^{,\alpha}=\frac{1}{r}\frac{\partial}{\partial\theta}Y\,,\qquad\phi_\alpha Y^{,\alpha}=\frac{1}{r\sin\theta}\frac{\partial}{\partial\phi}Y\equiv\frac{1}{r}\check\partial_\phi Y\,, \label{angularderivatives}
\end{align}
where we have defined $\check\partial_\phi\equiv\partial_\phi/\sin\theta$ for simplicity. Note that the operators $\partial_\theta$ and $\check\partial_\phi$ are not commuting, $\check\partial_\phi\partial_\theta\neq\partial_\theta\check\partial_\phi$. Furthermore, note that the non-vanishing derivatives of the unit vectors are given by
\begin{align}
&\frac{\partial}{\partial\theta}n_\alpha=\theta_\alpha\,,\qquad\frac{\partial}{\partial\phi}n_\alpha=\sin\theta\,\phi_\alpha\,,\qquad\frac{\partial}{\partial\theta}\theta_\alpha=-n_\alpha\,,\qquad\frac{\partial}{\partial\phi}\theta_\alpha=\cos\theta\,\phi_\alpha\,,\nonumber \\
&\frac{\partial}{\partial\theta}\phi_\alpha=0\,,\qquad\frac{\partial}{\partial\phi}\phi_\alpha=-\sin\theta\,n_\alpha-\cos\theta\,\theta_\alpha\,. \label{derivunitvectors}
\end{align}
\subsection{Diagonal components of the distortion matrix} \label{Subsection:D11D22}
Here we calculate the component $\check{\mathbb{D}}^2_2$ of the distortion matrix. The calculations of $\check{\mathbb{D}}^1_1$ are performed in a completely analogous way. First, recall that $\hat{\mathfrak{R}}^2_2$ is defined as:
\begin{align}
\hat{\mathfrak{R}}^2_2=&{\hat R^\alpha}{_{\mu\gamma\nu}}\hat k^\mu\hat k^\nu\phi_\alpha\phi^\gamma \nonumber \\
=&{\hat R^\alpha}{_{0\gamma 0}}\phi_\alpha\phi^\gamma+{\hat R^\alpha}{_{\beta\gamma \delta}}n^\beta n^\delta\phi_\alpha\phi^\gamma-{\hat R^\alpha}{_{\beta\gamma 0}}n^\beta\phi_\alpha\phi^\gamma-{\hat R^\alpha}{_{0\gamma\delta}}n^\delta\phi_\alpha\phi^\gamma \nonumber \\
=&{\hat R^\alpha}{_{0\gamma 0}}\phi_\alpha\phi^\gamma+{\hat R^\alpha}{_{\beta\gamma \delta}}n^\beta n^\delta\phi_\alpha\phi^\gamma-2{\hat R^\alpha}{_{\beta\gamma 0}}n^\beta\phi_\alpha\phi^\gamma \label{R22}\,.
\end{align}
We can now calculate $\hat{\mathfrak{R}}^2_2$ by inserting the expressions for the components of the Riemann tensor given in Section \ref{Subsection:Riemann} and by applying the properties of the spherical coordinates given in \eqref{angularderivatives} and \eqref{derivunitvectors}. The first and third terms on the right-hand side of equation \eqref{R22} are given by
\begin{align}
{\hat R^\alpha}{_{0\gamma 0}}\phi_\alpha\phi^\gamma
=&\left( {\mathcal{A}}^{,\alpha}{_\beta}-\frac{1}{2}\left({{\mathcal{B}}_\beta}{^{,\alpha}}+{{\mathcal{B}}^\alpha}{_{,\beta}}\right)'-{{\mathcal{C}}^\alpha_\beta}{''}\right)\phi_\alpha\phi^\gamma \nonumber \\
=&\frac{1}{\bar{r}^2}\check\partial_\phi^2{\mathcal{A}}+\frac{1}{\bar{r}}\partial_r{\mathcal{A}}+\frac{\cot\theta}{\bar{r}^2}\partial_\theta{\mathcal{A}}-\frac{1}{\bar{r}}\phi^\alpha\check\partial_\phi{\mathcal{B}}'_\alpha-\mathcal{C}^\alpha_{\gamma}{''}\phi_\alpha\phi^\gamma\label{1stterm}
\end{align}
and
\begin{align}
{\hat R^\alpha}{_{\beta\gamma 0}}n^\beta\phi_\alpha\phi^\gamma
=&\left(\frac{1}{2}\left({{\mathcal{B}}_\beta}{^{,\alpha}}-{{\mathcal{B}}^\alpha} {_{,\beta}}\right)_{,\gamma}-{\mathcal{C}}^\alpha_{\gamma,\beta}{'}+{{\mathcal{C}}{_{\beta\gamma}}}{^{,\alpha}}{'}\right)n^\beta\phi_\alpha\phi^\gamma \nonumber \\
=&\frac{1}{2\bar{r}^2}n^\beta \check\partial_\phi^2{\mathcal{B}}_{\beta}+\frac{1}{2\bar{r}}n^\beta \partial_{\bar r}{\mathcal{B}}_{\beta}+\frac{\cot\theta}{2\bar{r}^2}n^\beta \partial_\theta{\mathcal{B}}_{\beta}+\frac{1}{2\bar{r}^2}\phi^\alpha\check\partial_\phi{\mathcal{B}}_\alpha \nonumber \\
&-\frac{1}{2\bar{r}} \phi^\alpha\check\partial_\phi\partial_{\bar{r}}{\mathcal{B}}_{\alpha}-\phi^\alpha\phi^\gamma\partial_{\bar{r}}{\mathcal{C}}'_{\alpha\gamma}+\frac{1}{\bar{r}}n^\beta\phi^\gamma\check\partial_\phi{\mathcal{C}}'_{\beta\gamma}\,. \label{3rdterm}
\end{align}
To calculate the second term
\begin{align}
{\hat R^\alpha}{_{\beta\gamma \delta}}n^\beta n^\delta\phi_\alpha\phi^\gamma=&\left(2{\mathcal{C}}^\alpha{_{[\delta,\gamma]\beta}}+2{\mathcal{C}}_{\beta[\gamma,\delta]}{^\alpha}\right)\phi_\alpha\phi^\gamma n^\beta n^\delta \nonumber \\
=&\left(2{\mathcal{C}}^\alpha_{\beta,\delta\gamma}-{\mathcal{C}}^\alpha_{\gamma,\beta\delta}-{{\mathcal{C}}_{\beta\delta}}{^{,\alpha}}{_\gamma}\right)\phi_\alpha\phi^\gamma n^\beta n^\delta\,,
\end{align}
note that
\begin{align}
{{\mathcal{C}}^\alpha}{_{\beta,\delta\gamma}}\phi_\alpha\phi^\gamma n^\beta n^\delta=\frac{1}{\bar{r}}\phi_\alpha n^\beta n^\delta\partial_\phi{\mathcal{C}}^\alpha_{\beta,\delta}=\frac{1}{\bar{r}}\phi_\alpha n^\beta\check\partial_\phi\partial_{\bar{r}}{\mathcal{C}}^\alpha_{\beta}-\frac{1}{\bar{r}^2}\phi_\alpha n^\beta\check\partial_\phi\mathcal{C}^\alpha_{\beta}\,,
\end{align}
and that
\begin{align}
{{\mathcal{C}}_{\beta\delta}}{^{,\alpha}}{_\gamma}\phi_\alpha\phi^\gamma n^\beta n^\delta
=&\frac{1}{\bar{r}}n^\beta n^\delta\partial_\phi\left({{\mathcal{C}}_{\beta\delta}}{^{,\alpha}}\phi_\alpha\right)-\frac{1}{\bar{r}}{{\mathcal{C}}_{\beta\delta}}{^{,\alpha}}n^\beta n^\delta\check\partial_\phi\phi_\alpha \nonumber \\
=&\frac{1}{\bar{r}^2}n^\beta n^\delta\check\partial_\phi^2{{\mathcal{C}}_{\beta\delta}}+\frac{1}{\bar{r}}n^\beta n^\delta\partial_{\bar{r}}{{\mathcal{C}}_{\beta\delta}}+\frac{\cot\theta}{\bar{r}^2}n^\beta n^\delta \partial_\theta{{\mathcal{C}}_{\beta\delta}}\,.
\end{align}
Hence, we obtain for the second term on the right-hand side of equation \eqref{R22}:
\begin{align}
{\hat R^\alpha}{_{\beta\gamma \delta}}n^\beta n^\delta\phi_\alpha\phi^\gamma=&\frac{2}{\bar{r}}\phi_\alpha n^\beta \check\partial_\phi\partial_{\bar{r}}{\mathcal{C}}^\alpha_{\beta}-\frac{2}{\bar{r}^2}\phi_\alpha n^\beta \check\partial_\phi{\mathcal{C}}^\alpha_\beta-\phi_\alpha\phi^\gamma\partial_{\bar{r}}^2{\mathcal{C}}^\alpha_\gamma \nonumber \\
&-\frac{1}{\bar{r}^2}n^\beta n^\delta\check\partial_\phi^2{{\mathcal{C}}_{\beta\delta}}-\frac{1}{\bar{r}}n^\beta n^\delta \partial_{\bar{r}}{{\mathcal{C}}_{\beta\delta}}-\frac{\cot\theta}{\bar{r}^2}n^\beta n^\delta\partial_\theta{{\mathcal{C}}_{\beta\delta}}\,. \label{2ndterm2}
\end{align}
By summing up the three different terms, we obtain the following expression for the component $\hat{\mathfrak{R}}^2_2$ of the distortion matrix:
\begin{align}
\hat{\mathfrak{R}}^2_2=&\frac{1}{\bar{r}^2}\check\partial_\phi^2{\mathcal{A}}+\frac{1}{\bar{r}}\partial_{\bar{r}}{\mathcal{A}}+\frac{\cot\theta}{\bar{r}^2}\partial_\theta{\mathcal{A}}-\frac{2}{\bar{r}}\phi_\alpha n^\beta \frac{\mathrm d}{\mathrm d\lambda}\check\partial_\phi\mathcal{C}^\alpha_{\beta}-\frac{2}{\bar{r}^2}\phi_\alpha n^\beta \check\partial_\phi{\mathcal{C}}^\alpha_{\beta} \nonumber \\
&-\phi_\alpha\phi^\gamma\frac{\mathrm d^2}{\mathrm d\lambda^2}{\mathcal{C}}^\alpha_\gamma-\frac{1}{\bar{r}^2}n^\beta n^\delta\check\partial_\phi^2{{\mathcal{C}}_{\beta\delta}}-\frac{1}{\bar{r}}n^\beta n^\delta \partial_{\bar{r}}{{\mathcal{C}}_{\beta\delta}}-\frac{\cot\theta}{\bar{r}^2}n^\beta n^\delta \partial_\theta{{\mathcal{C}}_{\beta\delta}} \nonumber \\
&-\frac{1}{\bar{r}^2}n^\beta\check\partial_\phi^2{\mathcal{B}}_{\beta}-\frac{1}{\bar{r}}n^\beta \partial_{\bar{r}}{\mathcal{B}}_{\beta}-\frac{\cot\theta}{\bar{r}^2}n^\beta \partial_\theta{\mathcal{B}}_{\beta}-\frac{1}{\bar{r}}\phi^\alpha\frac{\mathrm d}{\mathrm d\lambda}\check\partial_\phi{\mathcal{B}}_\alpha-\frac{1}{\bar{r}^2}\phi^\alpha\check\partial_\phi{\mathcal{B}}_\alpha\,, \label{R22computed}
\end{align}
where we simplified the expression by using that
\begin{align}
\frac{\mathrm d^2}{\mathrm d\lambda^2}=\frac{\partial^2}{\partial\tau^2}-2\frac{\partial^2}{\partial\tau\partial\bar{r}}+\frac{\partial^2}{\partial\bar{r}^2}\,,
\end{align}
since $\mathrm d/\mathrm d\lambda=\partial_\tau-n^\alpha\partial_\alpha$. 
However, since we are concerned with the gauge transformation properties of the weak lensing observables, this expression for $\hat{\mathfrak{R}}^2_2$ in terms of $\mathcal{A}$, $\mathcal{B}^\alpha$ and $\mathcal{C}^{\alpha\beta}$ is impractical. Instead, we want to express it using the gauge-invariant quantities defined in equation~\eqref{gmetric}. For this, we first rewrite it in terms of the scalar, vector and tensor perturbations defined in equation~\eqref{MetricDecomposition}. 

First, consider the contribution of scalar perturbation $\alpha$, $\beta$, $\varphi$ and $\gamma$ to the quantity $\hat{\mathfrak{R}}^2_2$. By inserting $\mathcal{B}^\alpha=\beta^{,\alpha}+B^\alpha$ into \eqref{R22computed}, we obtain for the terms in $\beta$:
\begin{align}
\left(\hat{\mathfrak{R}}^2_2\right)_\beta=&-\frac{1}{\bar{r}^2}n^\alpha \check\partial_\phi^2\beta_{,\alpha}-\frac{1}{\bar{r}}n^\alpha \check\partial_{\bar{r}}\beta_{,\alpha}-\frac{\cot\theta}{\bar{r}^2}n^\beta {\partial_\theta}\beta_{,\alpha}-\frac{1}{\bar{r}}\phi^\alpha\frac{\mathrm d}{\mathrm d\lambda}\check\partial_\phi\beta_{,\alpha}-\frac{1}{\bar{r}^2}\phi^\alpha{\check\partial_\phi}\beta_{,\alpha}\,. \label{termsinbeta}
\end{align}
Using that
\begin{align}
n^\alpha{\check\partial_\phi^2}\beta_{,\alpha}={\check\partial_\phi^2}\partial_{\bar{r}}\beta-\beta_{,\alpha}{\check\partial_\phi^2}n^\beta-2{\check\partial_\phi}\beta_{,\alpha}{\check\partial_\phi}n^\alpha={\check\partial_\phi^2}\partial_{\bar{r}}\beta-\partial_{\bar{r}}\beta-\frac{\cot\theta}{\bar{r}}{\partial_\theta}\beta-2\frac{1}{\bar{r}}{\check\partial_\phi^2}\beta\,,
\end{align}
and that
\begin{align}
\frac{\mathrm d}{\mathrm d\lambda}\left(\phi^\alpha{\check\partial_\phi}\beta_{,\alpha}\right)=&\frac{\mathrm d}{\mathrm d\lambda}\left(\frac{1}{\bar{r}}{\check\partial_\phi^2}\beta+\partial_{\bar{r}}\beta+\frac{\cot\theta}{\bar{r}}{\partial_\theta}\beta\right) \nonumber \\
=&\frac{1}{\bar{r}}\frac{\mathrm d}{\mathrm d\lambda}{\check\partial_\phi^2}\beta+\frac{\mathrm d}{\mathrm d\lambda}\partial_{\bar{r}}\beta+\frac{\cot\theta}{\bar{r}}\frac{\mathrm d}{\mathrm d\lambda}{\partial_\theta}\beta+\frac{1}{\bar{r}^2}{\check\partial_\phi^2}\beta+\frac{\cot\theta}{\bar{r}^2}{\partial_\theta}\beta\,,
\end{align}
we can rewrite equation \eqref{termsinbeta} into
\begin{align}
\left(\hat{\mathfrak{R}}^2_2\right)_\beta=&-\frac{1}{\bar{r}^2}{\check\partial_\phi^2}\beta'-\frac{1}{\bar{r}}\partial_{\bar{r}}\beta'-\frac{\cot\theta}{\bar{r}^2}{\partial_\theta}\beta'\,.
\end{align}
By inserting $\mathcal{C}_{\alpha\beta}=\delta_{\alpha\beta}\varphi+\gamma_{,\alpha\beta}+C_{(\alpha,\beta)}+C_{\alpha\beta}$ into equation~\eqref{R22computed}, we obtain for the terms in $\gamma$:
\begin{align}
\left(\hat{\mathfrak{R}}^2_2\right)_\gamma=&-\frac{\mathrm d}{\mathrm d\lambda}\left(\frac{2}{\bar{r}}\phi^\alpha n^\beta {\check\partial_\phi}\gamma_{,\alpha\beta}\right)-\phi^\alpha\phi^\beta\frac{\mathrm d^2}{\mathrm d\lambda^2}\gamma_{,\alpha\beta} \nonumber \\
&-\frac{1}{\bar{r}^2} n^\beta n^\delta{\check\partial_\phi^2}\gamma_{,\beta\delta}-\frac{1}{\bar{r}} n^\beta n^\delta \partial_{\bar{r}}{\gamma_{,\beta\delta}}-\frac{\cot\theta}{\bar{r}^2}n^\beta n^\delta{\partial_\theta}{\gamma_{,\beta\delta}}\,. \label{R22gamma}
\end{align}
To simplify the first term of this expression, we use that
\begin{align}
\phi^\alpha n^\beta{\check\partial_\phi}\gamma_{,\alpha\beta}=&\partial_{\bar{r}}\left(\phi^\alpha{\check\partial_\phi}\partial_{\bar{r}}\gamma_{,\alpha}\right)-\frac{1}{\bar{r}}\phi^\alpha{\check\partial_\phi}\gamma_{,\alpha} \nonumber \\
=&\frac{1}{\bar{r}}{\check\partial_\phi^2}\partial_{\bar{r}}\gamma-\frac{2}{\bar{r}^2}{\check\partial_\phi^2}\gamma+\partial_{\bar{r}}^2\gamma+\frac{\cot\theta}{\bar{r}}{\partial_\theta}\partial_{\bar{r}}\gamma-\frac{2\cot\theta}{\bar{r}^2}{\partial_\theta}\gamma-\frac{1}{\bar{r}}\partial_{\bar{r}}\gamma\,. \label{termsGamma1}
\end{align}
For the third term, we use that
\begin{align}
{\check\partial_\phi^2}\left(n^\beta n^\delta\right)=-2 n^\beta n^\delta-\cot\theta\,\theta^\beta n^\delta-\cot\theta\,\theta^\delta n^\beta+2\phi^\beta\phi^\delta\,,
\end{align}
and, hence,
\begin{align}
n^\beta n^\delta{\check\partial_\phi^2}\gamma_{,\beta\delta}=&{\check\partial_\phi^2}\partial_{\bar{r}}^2\gamma-\gamma_{,\beta\delta}{\check\partial_\phi^2}\left(n^\beta n^\delta\right)-2{\check\partial_\phi}\gamma_{,\beta\delta}{\check\partial_\phi}\left(n^\beta n^\delta\right) \nonumber \\
=&{\check\partial_\phi^2}\partial_{\bar{r}}^2\gamma+\frac{6}{\bar{r}^2}{\check\partial_\phi^2}\gamma+\frac{2}{\bar{r}}\partial_{\bar{r}}\gamma-2\partial_{\bar{r}}^2\gamma-\frac{2\cot\theta}{\bar{r}}{\partial_\theta}\partial_{\bar{r}}\gamma-\frac{4}{\bar{r}}{\check\partial_\phi^2}\partial_{\bar{r}}\gamma+\frac{4\cot\theta}{\bar{r}^2}{\partial_\theta}\gamma\,. \label{termsGamma2}
\end{align}
By applying the relations given in equations \eqref{termsGamma1} and \eqref{termsGamma2} and also noting that
\begin{align}
\frac{\partial^2}{\partial\tau^2}=\frac{\partial^2}{\partial\bar{r}^2}+2\frac{\mathrm d}{\mathrm d\lambda}\frac{\partial}{\partial\bar{r}}+\frac{\mathrm d^2}{\mathrm d\lambda^2}\,,
\end{align}
we can show that equation \eqref{R22gamma} is equivalent to
\begin{align}
\left(\hat{\mathfrak{R}}^2_2\right)_\gamma=&-\frac{1}{\bar{r}^2}{\check\partial_\phi^2}\gamma''-\frac{1}{\bar{r}}\partial_{\bar{r}}\gamma''-\frac{\cot\theta}{\bar{r}^2}{\partial_\theta}\gamma''\,.
\end{align}
The expressions for the terms in $\alpha$ and $\varphi$ follow straight-forwardly from the expressions for $\hat{\mathfrak{R}}^2_2$. We obtain for the scalar terms in $\hat{\mathfrak{R}}^2_2$:
\begin{align}
\left(\hat{\mathfrak{R}}^2_2\right)_{s}=&\frac{1}{\bar{r}^2}{\check\partial_\phi^2}\left(\alpha-\varphi-\beta'-\gamma''\right)+\frac{1}{\bar{r}}\partial_{\bar{r}}\left(\alpha-\varphi-\beta'-\gamma''\right) \nonumber \\
&+\frac{\cot\theta}{\bar{r}^2}{\partial_\theta}\left(\alpha-\varphi-\beta'-\gamma''\right)-\frac{\mathrm d^2}{\mathrm d\lambda^2}\varphi\,.
\end{align}
Using the gauge-invariant variables defined in equation \eqref{gmetric}, we can rewrite this as
\begin{align}
\left(\hat{\mathfrak{R}}^2_2\right)_{s}=&\frac{1}{\bar{r}^2}{\check\partial_\phi^2}\left(\alpha_\chi-\varphi_\chi\right)-\frac{\mathrm d^2}{\mathrm d\lambda^2}\left(\alpha_\chi-\varphi_\chi\right)+\frac{1}{\bar{r}}\left(\alpha'_\chi-\varphi'_\chi\right) \nonumber \\
&+\frac{\cot\theta}{\bar{r}^2}{\partial_\theta}\left(\alpha_\chi-\varphi_\chi\right)-\frac{\mathrm d^2}{\mathrm d\lambda^2}\varphi_\chi-\frac{\mathrm d^2}{\mathrm d\lambda^2}\left(H\chi\right)\,.
\end{align}
Next we consider the vector perturbation $B_\alpha$ and $C_\alpha$. For the terms in $C_\alpha$, we obtain:
\begin{align}
\left(\hat{\mathfrak{R}}^2_2\right)_{C_\alpha}=&-\frac{2}{\bar{r}}\phi_\alpha n^\beta \frac{\mathrm d}{\mathrm d\lambda}{\check\partial_\phi}{C^\alpha}_{,\beta}-\frac{2}{\bar{r}^2}\phi_\alpha n^\beta {\check\partial_\phi}{C^\alpha}_{,\beta}-\phi^\alpha\phi^\gamma\frac{\mathrm d^2}{\mathrm d\lambda^2}C_{\alpha, \gamma} \nonumber \\
&-\frac{1}{\bar{r}^2} n^\beta n^\delta{\check\partial_\phi^2}{C_{\beta,\delta}}-\frac{1}{\bar{r}} n^\beta n^\delta \partial_{\bar{r}}{C_{\beta,\delta}}-\frac{\cot\theta}{\bar{r}^2}n^\beta n^\delta {\partial_\theta}{C_{\beta,\delta}}\,. \label{termsCalpha}
\end{align}
To simplify this expression, we need to rewrite all of the terms such that all derivatives are expressed in spherical coordinates. For example, for the fourth term we have:
\begin{align}
n^\beta n^\delta{ \check\partial_\phi^2}C_{\beta,\delta}
=&n^\beta{\check\partial_\phi^2}\partial_{\bar{r}}C_\beta-\frac{2}{\bar{r}}n^\beta{\check\partial_\phi^2} C_{\beta}-n^\beta\partial_{\bar{r}}C_{\beta}-\frac{\cot\theta}{\bar{r}}n^\beta{\partial_\theta}C_{\beta}\,.
\end{align}
Similar simplifications can be made for all other terms in the expression \eqref{termsCalpha}. We obtain:
\begin{align}
\left(\hat{\mathfrak{R}}^2_2\right)_{C_\alpha}=&-\frac{1}{\bar{r}}\phi_\alpha \frac{\mathrm d}{\mathrm d\lambda}{\check\partial_\phi}{C'^\alpha}-\frac{1}{\bar{r}^2}\phi_\alpha {\check\partial_\phi}{C'^\alpha}-\frac{1}{\bar{r}^2}n^\beta{\check\partial_\phi^2}C'_\beta-\frac{1}{\bar{r}} n^\beta \partial_{\bar{r}}C'_{\beta}-\frac{\cot\theta}{\bar{r}^2}n^\beta {\partial_\theta}C'_{\beta}\,.
\end{align}
Hence, the contributions of the vector perturbations $C_\alpha$ and $B_\alpha$ to $\hat{\mathfrak{R}}^2_2$ is given by
\begin{align}
\left(\hat{\mathfrak{R}}^2_2\right)_{v}=&-\frac{1}{\bar{r}}\phi_\alpha \frac{\mathrm d}{\mathrm d\lambda}{\check\partial_\phi}\left(B^\alpha+{C'^\alpha}\right)-\frac{1}{\bar{r}^2}\phi_\alpha {\check\partial_\phi}\left(B^\alpha+{C'^\alpha}\right)-\frac{1}{\bar{r}^2}n^\beta{\check\partial_\phi^2}\left(B_\beta+C'_\beta\right) \nonumber \\
&-\frac{1}{\bar{r}} n^\beta \partial_{\bar{r}}\left(B_\beta+C'_{\beta}\right)-\frac{\cot\theta}{\bar{r}^2}n^\beta {\partial_\theta}\left(B_\beta+C'_{\beta}\right)\,,
\end{align}
where the expression for the terms in $B_\alpha$ follows immediately from the expression \eqref{R22computed} for $\hat{\mathfrak{R}}^2_2$. Using the gauge-invariant quantity $\Psi^\alpha$ defined in \eqref{gmetric}, we can rewrite this as
\begin{align}
\left(\hat{\mathfrak{R}}^2_2\right)_{v}=&-\frac{1}{\bar{r}}\phi_\alpha \frac{\mathrm d}{\mathrm d\lambda}{\check\partial_\phi}\Psi^\alpha-\frac{1}{\bar{r}^2}\phi_\alpha {\check\partial_\phi}\Psi^\alpha-\frac{1}{\bar{r}^2}n^\beta{\check\partial_\phi^2}\Psi_\beta-\frac{1}{\bar{r}} n^\beta \partial_{\bar{r}}\Psi_\beta-\frac{\cot\theta}{\bar{r}^2}n^\beta {\partial_\theta}\Psi_\beta\,.
\end{align}
Summing up the expressions for the scalar, vector and tensor perturbations to $\hat{\mathfrak{R}}^2_2$, we obtain the following expression,
\begin{align}
\hat{\mathfrak{R}}^2_2=&\frac{1}{\bar{r}^2}{\check\partial_\phi^2}(\alpha_\chi-\varphi_\chi)-\frac{1}{\bar{r}}\frac{\mathrm d}{\mathrm d\lambda}(\alpha_\chi-\varphi_\chi)+\frac{1}{\bar{r}}(\alpha'_\chi-\varphi'_\chi)+\frac{\cot\theta}{\bar{r}^2}{\partial_\theta}(\alpha_\chi-\varphi_\chi)-\frac{\mathrm d^2}{\mathrm d\lambda^2}\varphi_\chi \nonumber \\
&-\frac{1}{\bar{r}^2}n^\beta{\check\partial_\phi^2}\Psi_\beta-\frac{1}{\bar{r}}n^\beta\partial_{\bar{r}}\Psi_\beta-\frac{\cot\theta}{\bar{r}^2}n^\beta{\partial_\theta}\Psi_\beta-\frac{1}{\bar{r}}\phi^\alpha\frac{\mathrm d}{\mathrm d\lambda}{\check\partial_\phi}\Psi_\alpha-\frac{1}{\bar{r}^2}\phi^\alpha{\check\partial_\phi}\Psi_\alpha \nonumber \\
&-\frac{2}{\bar{r}}\phi_\alpha n^\beta \frac{\mathrm d}{\mathrm d\lambda}{\check\partial_\phi}C^\alpha_{\beta}-\frac{2}{\bar{r}^2}\phi_\alpha n^\beta {\check\partial_\phi}C^\alpha_{\beta}-\phi_\alpha\phi^\gamma\frac{\mathrm d^2}{\mathrm d\lambda^2}C^\alpha_\gamma-\frac{1}{\bar{r}^2} n^\beta n^\delta{\check\partial_\phi^2}{C_{\beta\delta}} \nonumber \\
&-\frac{1}{\bar{r}} n^\beta n^\delta \partial_{\bar{r}}{C_{\beta\delta}}-\frac{\cot\theta}{\bar{r}^2}n^\beta n^\delta {\partial_\theta}{C_{\beta\delta}}-\frac{\mathrm d^2}{\mathrm d\lambda^2}(H\chi) \,, \label{R22result}
\end{align}
where the terms in $C_{\alpha\beta}$ follow directly from the expression \eqref{R22computed} for $\hat{\mathfrak{R}}^2_2$. The calculations for the component $\hat{\mathfrak{R}}^1_1$ can be performed completely analogously. It is given by:
\begin{align}
\hat{\mathfrak{R}}^1_1=&\frac{1}{\bar{r}^2}{\partial_\theta^2}(\alpha_\chi-\varphi_\chi)-\frac{1}{\bar{r}}\frac{\mathrm d}{\mathrm d\lambda}(\alpha_\chi-\varphi_\chi)+\frac{1}{\bar{r}}(\alpha'_\chi-\varphi'_\chi)-\frac{\mathrm d^2}{\mathrm d\lambda^2}\varphi_\chi-\frac{\mathrm d^2}{\mathrm d\lambda^2}(H\chi) \nonumber \\
&-\frac{1}{\bar{r}^2}n^\beta{\partial_\theta^2}\Psi_\beta-\frac{1}{\bar{r}}n^\beta\partial_{\bar{r}}\Psi_\beta-\frac{1}{\bar{r}}\theta^\alpha\frac{\mathrm d}{\mathrm d\lambda}{\partial_\theta}\Psi_\alpha-\frac{1}{\bar{r}^2}\theta^\alpha{\partial_\theta}\Psi_\alpha-\frac{2}{\bar{r}}\theta_\alpha n^\beta \frac{\mathrm d}{\mathrm d\lambda}{\partial_\theta}C^\alpha_\beta \nonumber \\
&-\frac{2}{\bar{r}^2}\theta_\alpha n^\beta {\partial_\theta}C^\alpha_{\beta}-\theta_\alpha\theta^\gamma\frac{\mathrm d^2}{\mathrm d\lambda^2}C^\alpha_\gamma-\frac{1}{\bar{r}^2} n^\beta n^\delta{\partial_\theta^2}{C_{\beta\delta}}-\frac{1}{\bar{r}} n^\beta n^\delta \partial_{\bar{r}}{C_{\beta\delta}} \,. \label{R11result}
\end{align}
Note that there is a gauge-dependent term, $\mathrm d^2/\mathrm d\lambda^2(H\chi)$, appearing in the expressions for $\hat{\mathfrak{R}}^1_1$ and $\hat{\mathfrak{R}}^2_2$. This term vanishes in the conformally transformed metric $\hat g_{\mu\nu}$ where $H=0$, but not in the full metric $g_{\mu\nu}$. Note that we introduced the conformally transformed metric to simplify the calculations along the photon geodesic, not because the physics would be invariant under the conformal transformation. Omitting the term $\mathrm d^2/\mathrm d\lambda^2(H\chi)$ would thus lead to an incorrect and gauge-dependent result for the trace of $\check{\mathbb{D}}^I_J$ as the physics described by the full metric $g_{\mu\nu}$ is not correctly accounted for. 

Now, we need to calculate the first-order distortion matrix which, as stated in equation~\eqref{solutionDIJ}, is given by:
\begin{align}
\frac{1}{\lambda_s}\delta\hat{\mathfrak{D}}^I_J=-\int_0^{\lambda_s}\mathrm d\lambda\,\left(\frac{\lambda_s-\lambda}{\lambda_s\lambda}\right)\lambda^2\hat{\mathfrak{R}}^I_J(\lambda)\,. \label{integral}
\end{align}
For this, we apply the following integrals:
\begin{align}
-\int_0^{\lambda_s}\mathrm d\lambda\,\left(\frac{\lambda_s-\lambda}{\lambda_s\lambda}\right)\lambda^2\left(\frac{1}{\bar{r}}Y\right)=&-\frac{1}{\bar{r}_z}\int_0^{\bar{r}_z}\mathrm d\bar{r}\,(\bar{r}_z-\bar{r})Y\,, \nonumber \\ 
-\int_0^{\lambda_s}\mathrm d\lambda\,\left(\frac{\lambda_s-\lambda}{\lambda_s\lambda}\right)\lambda^2\left(\frac{1}{\bar{r}}\frac{\mathrm d}{\mathrm d\lambda}Y\right)=&-Y_o+\frac{1}{\bar{r}_z}\int_0^{\bar{r}_z}\mathrm d\bar{r}\,Y\,, \nonumber \\
-\int_0^{\lambda_s}\mathrm d\lambda\,\left(\frac{\lambda_s-\lambda}{\lambda_s\lambda}\right)\lambda^2\left(\frac{\mathrm d^2}{\mathrm d\lambda^2}Y\right)=&-Y_s-Y_o+\frac{2}{\bar{r}_z}\int_0^{\bar{r}_z}\mathrm d\bar{r}\,Y\,, \label{standardI3}
\end{align}
where $Y$ is some generic scalar function. Any of the terms in equations \eqref{R22result} and \eqref{R11result} will take the form of one of the integrals in equation~\eqref{standardI3} when inserted into the integral in equation \eqref{integral}. We obtain the following expressions:
\begin{align}
\frac{1}{\lambda_s}\delta\hat{\mathfrak{D}}^2_2=&\left(\alpha_\chi+H\chi+\phi^\alpha\phi^\beta C_{\alpha\beta}\right)_o+\left(\varphi_\chi+H\chi+\phi^\alpha\phi^\beta C_{\alpha\beta}\right)_s \nonumber \\
&+\int_0^{\lambda_s}\mathrm d\lambda\left(\frac{\lambda_s-\lambda}{\lambda_s\lambda}\right)\left[-\left(\cot\theta\,\partial_\theta+\check\partial^2_\phi\right)(\alpha_\chi-\varphi_\chi)+\left(n^\alpha\check\partial^2_\phi+\cot\theta\, n^\alpha\partial_\theta+\phi^\alpha\check\partial_\phi\right)\Psi_\alpha\right] \nonumber \\
&+\int_0^{\lambda_s}\mathrm d\lambda\left(\frac{\lambda_s-\lambda}{\lambda_s\lambda}\right)\left(2\phi^\alpha n^\beta\check\partial_\phi+n^\alpha n^\beta\check\partial^2_\phi+\cot\theta\,n^\alpha n^\beta\partial_\theta\right)C_{\alpha\beta} \nonumber \\
&-\frac{1}{\bar{r}_z}\int_0^{\bar{r}_z}\mathrm d\bar{r}\,\left(\alpha_\chi-\varphi_\chi+2\varphi_\chi+2H\chi+\phi^\alpha\check\partial_\phi\Psi_\alpha+2\phi_\alpha n^\beta\check\partial_\phi C^\alpha_\beta+2\phi^\alpha\phi^\beta C_{\alpha\beta}\right) \nonumber \\
&+\frac{1}{\bar{r}_z}\int_0^{\bar{r}_z}\mathrm d\bar{r}\,\left(\bar{r}_z-\bar{r}\right)\left(\varphi'_\chi-\alpha'_\chi+n^\alpha\partial_{\bar{r}}\Psi_\alpha+n^\alpha n^\beta\partial_{\bar{r}}C_{\alpha\beta}\right)\,,
\end{align}
and
\begin{align}
\frac{1}{\lambda_s}\delta\hat{\mathfrak{D}}^1_1=&\left(\alpha_\chi+H\chi+\theta^\alpha\theta^\beta C_{\alpha\beta}\right)_o+\left(\varphi_\chi+H\chi+\theta^\alpha\theta^\beta C_{\alpha\beta}\right)_s \nonumber \\
&+\int_0^{\lambda_s}\mathrm d\lambda\left(\frac{\lambda_s-\lambda}{\lambda_s\lambda}\right)\left[-\partial^2_\theta(\alpha_\chi-\varphi_\chi)+\left(n^\alpha\partial^2_\theta+\theta^\alpha\partial_\theta\right)\Psi_\alpha\right] \nonumber \\
&+\int_0^{\lambda_s}\mathrm d\lambda\left(\frac{\lambda_s-\lambda}{\lambda_s\lambda}\right)\left(2\theta^\alpha n^\beta\partial_\theta+n^\alpha n^\beta\partial^2_\theta\right)C_{\alpha\beta} \nonumber \\
&-\frac{1}{\bar{r}_z}\int_0^{\bar{r}_z}\mathrm d\bar{r}\,\left(\alpha_\chi-\varphi_\chi+2\varphi_\chi+2H\chi+\theta^\alpha\partial_\theta\Psi_\alpha+2\theta_\alpha n^\beta\partial_\theta C^\alpha_\beta+2\theta^\alpha\theta^\beta C_{\alpha\beta}\right) \nonumber \\
&+\frac{1}{\bar{r}_z}\int_0^{\bar{r}_z}\mathrm d\bar{r}\,\left(\bar{r}_z-\bar{r}\right)\left(\varphi'_\chi-\alpha'_\chi+n^\alpha\partial_{\bar{r}}\Psi_\alpha+n^\alpha n^\beta\partial_{\bar{r}}C_{\alpha\beta}\right)\,.
\end{align}
Following equation \eqref{DeformationMatrix}, the components $\check{\mathbb{D}}^1_1$ and $\check{\mathbb{D}}^2_2$ of the distortion matrix are now determined by
\begin{align}
\check{\mathbb{D}}^1_1=1+\delta z+\widehat{\Delta\nu}_o-\frac{\Delta\lambda_z}{\bar r_z}+\frac{1}{\lambda_s}\delta\hat{\mathfrak{D}}^1_1\,,\qquad\check{\mathbb{D}}^2_2=1+\delta z+\widehat{\Delta\nu}_o-\frac{\Delta\lambda_z}{\bar r_z}+\frac{1}{\lambda_s}\delta\hat{\mathfrak{D}}^2_2\,. \label{D11D22}
\end{align}
\subsection{Off-diagonal components of the distortion matrix} \label{Subsection:D12}
Here we calculate the off-diagonal components $\check{\mathbb{D}}^1_2$ and $\check{\mathbb{D}}^2_1$ of the distortion matrix. For this, we first calculate the source term $\hat{\mathfrak{R}}^I_J$, which is 
\begin{align}
\hat{\mathfrak{R}}^1_2=\hat{\mathfrak{R}}^2_1\equiv&{\hat R^\alpha}{_{\mu\gamma \nu}}\hat k^\mu\hat k^\nu\theta_\alpha\phi^\gamma \nonumber \\
=&{\hat R^\alpha}{_{0\gamma 0}}\theta_\alpha\phi^\gamma+{\hat R^\alpha}{_{\beta\gamma \delta}}n^\beta n^\delta\theta_\alpha\phi^\gamma-{\hat R^\alpha}{_{\beta\gamma 0}}n^\beta\theta_\alpha\phi^\gamma-{\hat R^\alpha}{_{0\gamma\delta}}n^\delta\theta_\alpha \phi^\gamma\,. \label{R12Def}
\end{align}
With the same techniques as in the calculation of $\hat{\mathfrak{R}}^2_2$, we will now compute the four different terms. To abbreviate the lengthy expressions in this section, we introduce the differential operator
\begin{align}
\mathrm{X}^\alpha\equiv\theta^\alpha\frac{1}{\sin\theta}\frac{\partial}{\partial\phi}+\phi^\alpha\frac{\partial}{\partial\theta}\,,
\end{align}
Inserting the expressions of the Riemann tensor given in Section~\ref{Subsection:Riemann}, we obtain
\begin{align}
{\hat R^\alpha}{_{0\gamma 0}}\theta_\alpha\phi^\gamma
=&\left({\mathcal{A}}^{,\alpha}{_\beta}-\frac{1}{2}\left({{\mathcal{B}}_\beta}{^{,\alpha}}+{{\mathcal{B}}^\alpha}{_{,\beta}}\right)'-{{\mathcal{C}}^\alpha_\beta}{''}\right)\theta_\alpha\phi^\gamma \nonumber \\
=&\frac{1}{\bar{r}^2}\partial_\theta\check\partial_\phi{\mathcal{A}}-\frac{1}{2\bar{r}}\mathrm{X}_\alpha{{\mathcal{B}}^{\alpha}}'-{\mathcal{C}}_{\alpha\gamma}''\theta^\alpha\phi^\gamma \label{1sttermR12}
\end{align}
for the first term on the right-hand side of equation \eqref{R12Def},
\begin{align}
{\hat R^\alpha}{_{\beta\gamma \delta}}n^\beta n^\delta\theta_\alpha\phi^\gamma=&\left(2{\mathcal{C}}^\alpha{_{(\beta,\delta)\gamma}}-2{\mathcal{C}}^\alpha{_{(\beta,\gamma)\delta}}+{{\mathcal{C}}_{\beta\gamma}}{^{,\alpha}}{_\delta}-{{\mathcal{C}}_{\beta\delta}}{^{,\alpha}}{_\gamma}\right)\theta_\alpha\phi^\gamma n^\beta n^\delta \nonumber \\
=&\left({{\mathcal{C}}^\alpha}{_{\beta,\delta\gamma}}+{\mathcal{C}_{\beta\gamma}}{^{,\alpha}}{_\delta}\right)\theta_\alpha\phi^\gamma n^\beta n^\delta-\left({{\mathcal{C}}^\alpha}{_{\gamma,\beta\delta}}+{{\mathcal{C}}_{\beta\delta}}{^{,\alpha}}{_\gamma}\right)\theta_\alpha\phi^\gamma n^\beta n^\delta \nonumber \\
=&\frac{1}{\bar{r}}n^\beta\mathrm X^\alpha\partial_{\bar{r}}{\mathcal{C}}_{\alpha\beta}-\frac{1}{\bar{r}^2}n^\beta\mathrm X^\alpha{\mathcal{C}}{_{\alpha\beta}}-\theta^\alpha\phi^\gamma\partial^2_{\bar{r}}{\mathcal{C}}{_{\alpha\gamma}}-\frac{1}{\bar{r}^2}n^\beta n^\gamma\partial_\theta\check\partial_\phi{\mathcal{C}}{_{\beta\gamma}}
\end{align}
for the second term,
\begin{align}
{\hat R^\alpha}{_{\beta\gamma 0}}n^\beta\theta_\alpha\phi^\gamma
=&\left(\frac{1}{2}\left({{\mathcal{B}}_\beta}{^{,\alpha}}-{{\mathcal{B}}^\alpha} {_{,\beta}}\right)_{,\gamma}-\mathcal{C}^\alpha_{\gamma,\beta}{'}+{{\mathcal{C}}{_{\beta\gamma}}}{^{,\alpha}}{'}\right)n^\beta\theta_\alpha\phi^\gamma \nonumber \\
=&\frac{1}{2\bar{r}^2}n^\beta\partial_\theta\check\partial_\phi{\mathcal{B}}_\beta-\frac{1}{2\bar{r}}\theta_\alpha\check\partial_\phi\partial_{\bar{r}}{\mathcal{B}}^\alpha+\frac{1}{2\bar{r}^2}\theta_\alpha\check\partial_\phi{\mathcal{B}}^\alpha \nonumber \\
&-\theta^\alpha\phi^\gamma\partial_{\bar{r}}{\mathcal{C}}_{\alpha\gamma}'+\frac{1}{\bar{r}}n^\beta\phi^\gamma{\partial_\theta}{\mathcal{C}}'_{\beta\gamma}
\end{align}
for the third term, and
\begin{align}
\hat R^\alpha{_{0\gamma\delta}}n^\delta \theta_\alpha\phi^\gamma=&\left(-{{\mathcal{B}}_{[\beta}}{^{,\alpha}}{_{\gamma]}}-2{{\mathcal{C}}^\alpha}{_{[\beta,\gamma]}}'\right)n^\delta \theta_\alpha\phi^\gamma \nonumber \\
=&-\frac{1}{2\bar{r}}\phi^\gamma{\partial_\theta}\partial_{\bar{r}}{\mathcal{B}}_\gamma+\frac{1}{2\bar{r}^2}\phi^\gamma{\partial_\theta}{\mathcal{B}}_\gamma+\frac{1}{2\bar{r}^2}n^\delta\partial_\theta\check\partial_\phi{\mathcal{B}}_\delta \nonumber \\
&-\theta^\alpha\phi^\gamma\partial_{\bar{r}}{\mathcal{C}}'_{\alpha\gamma}+\frac{1}{\bar{r}}\theta^\alpha n^\delta {\check\partial_\phi}{\mathcal{C}}'_{\alpha\delta} \label{4thtermR12}
\end{align}
for the fourth term. Summing all these four terms, we obtain for $\hat{\mathfrak{R}}^1_2$:
\begin{align}
\hat{\mathfrak{R}}^1_2=&\frac{1}{\bar{r}^2}\partial_\theta\check\partial_\phi{\mathcal{A}}-\frac{1}{2\bar{r}}\mathrm X_\alpha{\mathcal{B}}^{\alpha'}+\frac{1}{2\bar{r}}\mathrm X_\alpha\frac{\partial}{\partial\bar{r}}{\mathcal{B}}^\alpha-\frac{1}{2\bar{r}^2}\mathrm X_\alpha{\mathcal{B}}^\alpha-\frac{1}{\bar{r}^2}n^\alpha\partial_\theta\check\partial_\phi{\mathcal{B}}_\alpha\nonumber \\
&-\theta^\alpha\phi^\gamma\frac{\mathrm d^2}{\mathrm d\lambda^2}{\mathcal{C}}_{\alpha\gamma}-\frac{1}{\bar{r}}n^\beta\mathrm{X}^\alpha\frac{\mathrm d}{\mathrm d\lambda}{\mathcal{C}}_{\alpha\beta}-\frac{1}{\bar{r}^2}n^\beta\mathrm X^\alpha{\mathcal{C}}_{\alpha\beta}-n^\beta n^\gamma\frac{1}{\bar{r}^2}\partial_\theta\check\partial_\phi{\mathcal{C}}_{\beta\gamma}\,. \label{R12}
\end{align}
We now want to express this using the metric decompositions given in \eqref{MetricDecomposition}. As for the calculations of $\hat{\mathfrak{R}}^2_2$, we need to do some work to simplify the expressions for the terms in $\beta$, $\gamma$ and $C^\alpha$. For this, we apply the following relations:
\begin{align}
&\mathrm X_\alpha\nabla^\alpha=\frac{1}{\bar{r}}\mathrm X_\alpha\widehat\nabla^\alpha=\frac{2}{\bar{r}}\partial_\theta\check\partial_\phi\,,\qquad \mathrm X_\beta n^\beta=\theta^\alpha\phi^\beta+\theta^\beta\phi^\alpha\,,\nonumber \\
&\partial_\theta\check\partial_\phi n^\alpha=0\,,\qquad\partial_\theta\check\partial_\phi\left(n^\alpha n^\beta\right)=\theta^\alpha\phi^\beta+\theta^\beta\phi^\alpha\,. \label{helpfulrelations}
\end{align}
First, we consider the scalar contributions to $\hat{\mathfrak{R}}^1_2$. By inserting $\mathcal{B}^\alpha=\beta^{,\alpha}+B^\alpha$ into the equation~\eqref{R12}, we obtain for the terms in $\beta$:
\begin{align}
\left({\hat{\mathfrak{R}}^1_2}\right)_\beta=&-\frac{1}{2\bar{r}}\mathrm X_\alpha{\beta^{,\alpha}}'+\frac{1}{2\bar{r}}\mathrm X_\alpha\frac{\partial}{\partial\bar{r}}\beta^{,\alpha}-\frac{1}{2\bar{r}^2}\mathrm X_\alpha\beta^{,\alpha}-\frac{1}{\bar{r}^2}n^\alpha\partial_\theta\check\partial_\phi\beta^{,\alpha}=-\frac{1}{\bar{r}^2}\partial_\theta\check\partial_\phi\beta'\,,
\end{align}
where the second equality follows straight-forwardly from the relations \eqref{helpfulrelations}.
Using $\mathcal{C}^{\alpha\beta}=\delta^{\alpha\beta}\varphi+\gamma^{,\alpha\beta}+C^{(\alpha,\beta)}+C^{\alpha\beta}$, we obtain for the terms in $\gamma$:
\begin{align}
\left({\hat{\mathfrak{R}}^1_2}\right)_\gamma=&-\theta^\alpha\phi^\gamma\frac{\mathrm d^2}{\mathrm d\lambda^2}\gamma_{,\alpha\gamma}-\frac{1}{\bar{r}}n^\beta\mathrm X^\alpha\frac{\mathrm d}{\mathrm d\lambda}\gamma_{,\alpha\beta}-\frac{1}{\bar{r}^2}n^\beta\mathrm X^\alpha\gamma_{,\alpha\beta}-n^\beta n^\gamma\frac{1}{\bar{r}^2}\partial_\theta\check\partial_\phi\gamma_{,\beta\gamma} \nonumber \\
=&-\frac{1}{\bar{r}^2}\partial_\theta\check\partial_\phi\gamma''\,, 
\end{align}
where for the second equality we used that
\begin{align}
n^\beta\mathrm X^\alpha\gamma_{,\alpha\beta}=&\frac{\partial}{\partial\bar{r}}\left(\frac{2}{\bar{r}}\partial_\theta\check\partial_\phi\gamma\right)-\gamma_{,\alpha\beta}\mathrm X^\alpha n^\beta=\frac{2}{\bar{r}}\frac{\partial}{\partial\bar{r}}\partial_\theta\check\partial_\phi\gamma-\frac{4}{\bar{r}^2}\partial_\theta\check\partial_\phi\gamma\,,
\end{align}
and that
\begin{align}
n^\beta n^\gamma\partial_\theta\check\partial_\phi\gamma_{,\beta\gamma}=&\partial_\theta\check\partial_\phi\left(n^\beta n^\gamma\gamma_{,\beta\gamma}\right)-\partial_\theta\check\partial_\phi\left(n^\beta n^\gamma\right)\gamma_{,\beta\gamma}-2n^\beta\left(\theta_\alpha\check\partial_\phi+\phi_\alpha\partial_\theta\right)\gamma_{,\beta\gamma} \nonumber \\
=&\partial_\theta\check\partial_\phi\frac{\partial^2}{\partial\bar{r}^2}\gamma+\frac{6}{\bar{r}^2}\partial_\theta\check\partial_\phi\gamma-\frac{4}{\bar{r}}\frac{\partial}{\partial\bar{r}}\partial_\theta\check\partial_\phi\gamma\,.
\end{align}
The terms in $\alpha$ and $\varphi$ follow straight-forwardly from the expression \eqref{R12}. For the contribution of all scalar perturbations to $\hat{\mathfrak{R}}^1_2$, we obtain:
\begin{align}
\left(\hat{\mathfrak{R}}^1_2\right)_s=\frac{1}{\bar{r}^2}\partial_\theta\check\partial_\phi\left(\alpha-\varphi-\beta'-\gamma''\right)=\frac{1}{\bar{r}^2}\left(\alpha_\chi-\varphi_\chi\right)\,.
\end{align}
Now, we consider the vector contributions to $\hat{\mathfrak{R}}^1_2$. For the terms in $C^\alpha$, we have
\begin{align}
\left({\hat{\mathfrak{R}}^1_2}\right)_{C^\alpha}=&-\theta^\alpha\phi^\gamma\frac{\mathrm d^2}{\mathrm d\lambda^2}C_{(\alpha,\gamma)}-\frac{1}{\bar{r}}n^\beta\mathrm X^\alpha\frac{\mathrm d}{\mathrm d\lambda}C_{(\alpha,\beta)}-\frac{1}{\bar{r}^2}n^\beta\mathrm X^\alpha C_{(\alpha,\beta)} \nonumber \\
&-n^\beta n^\gamma\frac{1}{\bar{r}^2}\partial_\theta\check\partial_\phi C_{(\beta,\gamma)}\,. 
\end{align} 
By applying the relations
\begin{align}
\theta^\alpha\phi^\gamma\frac{\mathrm d^2}{\mathrm d\lambda^2}C_{(\alpha,\gamma)}=&\theta^{(\alpha}\phi^{\gamma)}\frac{\mathrm d^2}{\mathrm d\lambda^2}C_{\alpha,\gamma}=\frac{1}{2}\frac{\mathrm d^2}{\mathrm d\lambda^2}\left(\frac{1}{\bar{r}}\mathrm X^\alpha C_\alpha\right) \nonumber \\
=&\frac{1}{2\bar{r}}\frac{\mathrm d^2}{\mathrm d\lambda^2}\left(\mathrm X^\alpha C_\alpha\right)+\frac{1}{\bar{r}^2}\frac{\mathrm d}{\mathrm d\lambda}\mathrm X^\alpha C_\alpha+\frac{1}{\bar{r}^3}\mathrm X^\alpha C_\alpha\,, 
\end{align}
and
\begin{align}
\frac{\mathrm d}{\mathrm d\lambda}\left(n^\beta\mathrm{X}^\alpha C_{\alpha,\beta}\right)=&\frac{\mathrm d}{\mathrm d\lambda}\frac{\partial}{\partial\bar{r}}\left(\mathrm{X}^\alpha C_\alpha\right)-\frac{\mathrm d}{\mathrm d\lambda}\left(C_{\alpha,\beta}\mathrm{X}^\alpha n^\beta\right) \nonumber \\
=&-\frac{\mathrm d^2}{\mathrm d\lambda^2}\left(\mathrm X^\alpha C_\alpha\right)+\frac{\mathrm d}{\mathrm d\lambda}\left(\mathrm X^\alpha C'_\alpha\right)-\frac{1}{\bar{r}}\frac{\mathrm d}{\mathrm d\lambda}\left(\mathrm X^\alpha C_\alpha\right)-\frac{1}{\bar{r}^2}\left(\mathrm X^\alpha C_\alpha\right)\,, 
\end{align}
and
\begin{align}
\partial_\theta\check\partial_\phi\left(n^\gamma C_{\beta,\gamma}\right)=&n^\gamma\partial_\theta\check\partial_\phi C_{\beta,\gamma}+\mathrm X^\gamma C_{\beta,\gamma}=n^\gamma\partial_\theta\check\partial_\phi C_{\beta,\gamma}+\frac{2}{\bar{r}}\partial_\theta\check\partial_\phi C_\beta\,,
\end{align}
we can simplify the expression for the terms in $C^\alpha$ to:
\begin{align}
\left(\hat{\mathfrak{R}}^1_2\right)_{C_\alpha}=&-\frac{1}{2\bar{r}}\frac{\mathrm d}{\mathrm d\lambda}\left(\mathrm X^\alpha C'_\alpha\right)-\frac{1}{2\bar{r}^2}\mathrm X^\alpha C'_\alpha-\frac{1}{\bar{r}^2}n^\beta\partial_\theta\check\partial_\phi C'_\beta\,.
\end{align}
Combining this with the terms in $B^\alpha$ which are obtained from expression \eqref{R12} without further simplifications, the total contribution of vector perturbations to $\hat{\mathfrak{R}}^1_2$ are given by:
\begin{align}
\left(\hat{\mathfrak{R}}^1_2\right)_v=&-\frac{1}{2\bar{r}}\frac{\mathrm d}{\mathrm d\lambda}\left(\mathrm X^\alpha \Psi_\alpha\right)-\frac{1}{2\bar{r}^2}\mathrm X^\alpha \Psi_\alpha-\frac{1}{\bar{r}^2}n^\beta\partial_\theta\check\partial_\phi\Psi_\beta\,.
\end{align}
Summing up the scalar, the vector and also the tensor contributions, we obtain for $\hat{\mathfrak{R}}^1_2$:
\begin{align}
\hat{\mathfrak{R}}^1_2=&\frac{1}{\bar{r}^2}\partial_\theta\check\partial_\phi\left(\alpha_\chi-\varphi_\chi\right)-\frac{1}{2\bar{r}}\frac{\mathrm d}{\mathrm d\lambda}\mathrm X^\alpha\Psi_\alpha-\frac{1}{2\bar{r}^2}\mathrm X^\alpha\Psi_\alpha-\frac{1}{\bar{r}^2}n^\alpha\partial_\theta\check\partial_\phi\Psi_\alpha \nonumber \\
&-\theta^\alpha\phi^\gamma\frac{\mathrm d^2}{\mathrm d\lambda^2}C_{\alpha\gamma}-\frac{1}{\bar{r}}n^\beta\mathrm X^\alpha\frac{\mathrm d}{\mathrm d\lambda}C_{\alpha\beta}-\frac{1}{\bar{r}^2}n^\beta\mathrm X^\alpha C_{\alpha\beta}-n^\beta n^\gamma\frac{1}{\bar{r}^2}\partial_\theta\check\partial_\phi C_{\beta\gamma}\,.
\end{align}
Now, we can calculate the components $\check{\mathbb{D}}^1_2=\check{\mathbb{D}}^2_1$ of the distortion matrix by performing the integration, applying the integrals given in equation~\eqref{standardI3}, which yields
\begin{align}
\check{\mathbb{D}}^1_2=&\left(\theta^\alpha\phi^\beta C_{\alpha\beta}\right)_o+\left(\theta^\alpha\phi^\beta C_{\alpha\beta}\right)_s-\frac{1}{2\bar{r}_z}\int_0^{\bar{r}_z}\mathrm d\bar{r}\,\mathrm X^\alpha\left(\Psi_\alpha+2n^\beta C_{\alpha\beta}\right)  \nonumber \\
&-\int_0^{\bar{r}_z}\mathrm d\bar{r}\,\left(\frac{\bar{r}_z-\bar{r}}{\bar{r}_z\bar{r}}\right)\left(\partial_\theta\check\partial_\phi(\alpha_\chi-\varphi_\chi)-n^\alpha\partial_\theta\check\partial_\phi\Psi_\alpha-n^\beta n^\gamma\partial_\theta\check\partial_\phi C_{\beta\gamma}\right) \nonumber \\
&+\int_0^{\bar{r}_z}\mathrm d\bar{r}\,\left(\frac{\bar{r}_z-\bar{r}}{\bar{r}_z\bar{r}}\right)\left(\frac{1}{2}\mathrm X^\alpha\Psi_\alpha+n^\beta\mathrm X^\alpha C_{\alpha\beta}\right)\,. \label{D12result}
\end{align} 
By applying the relations
\begin{align}
&\partial_\theta\check\partial_\phi\left(n^\alpha\Psi_\alpha\right)=n^\alpha\partial_\theta\check\partial_\phi\Psi_\alpha+\mathrm X_\alpha\Psi^\alpha\,,\nonumber \\
&\partial_\theta\check\partial_\phi\left(n^\alpha n^\beta C_{\alpha\beta}\right)=n^\alpha n^\beta\partial_\theta\check\partial_\phi C_{\alpha\beta}+2n^\alpha\mathrm X^\beta C_{\alpha\beta}+C_{\alpha\beta}\mathrm X^\alpha n^\beta\,,
\end{align}
the expression for $\check{\mathbb{D}}^1_2$ can be rewritten into
\begin{align}
\check{\mathbb{D}}^1_2=&\left(\theta^\alpha\phi^\beta C_{\alpha\beta}\right)_o+\left(\theta^\alpha\phi^\beta C_{\alpha\beta}\right)_s-\int_0^{\bar{r}_z}\frac{\mathrm d\bar{r}}{2\bar{r}}\,\mathrm X^\alpha\left(\Psi_\alpha+2n^\beta C_{\alpha\beta}\right)  \nonumber \\
&-\int_0^{\bar{r}_z}\mathrm d\bar{r}\,\left(\frac{\bar{r}_z-\bar{r}}{\bar{r}_z\bar{r}}\right)\partial_\theta\check\partial_\phi\left(\alpha_\chi-\varphi_\chi-n^\alpha\Psi_\alpha-n^\beta n^\gamma C_{\beta\gamma}\right)\,, \label{D12result}
\end{align} 
which, up to a negative sign, is equal to the expression \eqref{gamma2JMsolved} for the shear component $\check\gamma_2$.
\section{Gauge-Invariant Expression for the Distortion in the Luminosity Distance} \label{Appendix:LumDist}
Equation~\eqref{D11D22} for $\check{\mathbb{D}}^1_1$ and $\check{\mathbb{D}}^2_2$ yields, combined with the equation~\eqref{kappaJM} for $\delta D=-\check\kappa$, the following expression for $\check\kappa$:
\begin{align}
\check\kappa=&-\delta z-\widehat{\Delta\nu}_o+\frac{\Delta\lambda_s}{\bar{r}_z}-\left(\alpha_\chi+H\chi-\frac{1}{2}n^\alpha n^\beta C_{\alpha\beta}\right)_o-\left(\varphi_\chi+H\chi-\frac{1}{2}n^\alpha n^\beta C_{\alpha\beta}\right)_s \nonumber \\
&+\int_0^{\bar{r}_z}\mathrm d\bar{r}\,\left(\frac{\bar{r}_z-\bar{r}}{2\bar{r}_z\bar{r}}\right)\left(\widehat\nabla^2(\alpha_\chi-\varphi_\chi)-n^\alpha\widehat\nabla^2\Psi_\alpha-n^\alpha n^\beta\widehat\nabla^2 C_{\alpha\beta}\right) \nonumber \\
&+\frac{1}{\bar{r}_z}\int_0^{\bar{r}_z}\mathrm d\bar{r}\,\left(\alpha_\chi+\varphi_\chi+2H\chi+\frac{1}{2}\widehat\nabla^\alpha\Psi_\alpha+n^\beta\widehat\nabla^\alpha C_{\alpha\beta}-n^\alpha n^\beta C_{\alpha\beta}\right) \nonumber \\
&-\frac{1}{\bar{r}_z}\int_0^{\bar{r}_z}\mathrm d\bar{r}\,(\bar{r}_z-\bar{r})\left(\varphi'_\chi-\alpha'_\chi\right)\,. \label{kappaJMresult}
\end{align}
Due to the contributions of the perturbation quantities $\delta z$, $\widehat{\Delta\nu}_o$ and $\Delta\lambda_s$, the gauge-transformation property of this expression is not immediately evident. To deal with these terms, we relate the distortion $\Delta\lambda_s$ of the affine parameter to the distortion $\delta r$ of the radial coordinate. First, note that integrating the spatial part of the equation $\mathrm dx^\alpha/\mathrm d\lambda=\hat k^\alpha$ yields
\begin{align}
x^\alpha_s=\int_0^{\lambda_z+\Delta\lambda_s}\mathrm d\lambda\,\left(-n^\alpha-\delta n^\alpha\right)=\delta x^\alpha_o+\bar{r}_z-\Delta\lambda_s+\int_0^{\bar{r}_z}\mathrm d\lambda\,\delta n^\alpha\,,
\end{align}
which means that for $\delta r=x^\alpha_sn_\alpha-\bar{r}_z$, we have
\begin{align}
\delta r=\delta x^\alpha_on_\alpha-\Delta\lambda_s+\int_0^{\bar{r}_z}\mathrm d\bar{r}\,n_\alpha\delta n^\alpha\,.
\end{align}
Note that the photon wavevector fulfills the null condition $\hat k^\mu\hat k_\mu=0$, which, to first order, reads:
\begin{align}
0=n_\alpha\delta n^\alpha-\delta\nu-\mathcal{A}+\mathcal{B}_\alpha n^\alpha+\mathcal{C}_{\alpha\beta}n^\alpha n^\beta\,.
\end{align}
This enables us to write the equation for $\delta r$ as
\begin{align}
\delta r=\delta x^\alpha_on_\alpha-\Delta\lambda_s+\int_0^{\bar{r}_z}\mathrm d\bar{r}\,\delta\nu+\int_0^{\bar{r}_z}\mathrm d\bar{r}\,\left(\mathcal{A}-\mathcal{B}_\alpha n^\alpha-\mathcal{C}_{\alpha\beta}n^\alpha n^\beta\right)\,.
\end{align}
Now, we want to express the integral over $\delta\nu$ in terms of metric perturbations, for which we apply the temporal part of the geodesic equation,
\begin{align}
\frac{\mathrm d\hat k^a}{\mathrm d\lambda}=-\hat\Gamma^a{_{bc}}\hat k^b\hat k^c=-\hat\Gamma^a{_{bc}}\hat{\bar{k}}^b\hat{\bar{k}}^c\,,
\end{align}
where for the second equality we used that the Christoffel symbols of the conformally transformed metric vanish in the background. The equation for $\delta r$ now reads
\begin{align}
\delta r=\delta x^\alpha_on_\alpha-\Delta\lambda_s+\bar{r}_z\delta\nu_o+\int_0^{\bar{r}_z}\mathrm d\bar{r}\,(\bar{r}_z-\bar{r})\delta\hat\Gamma^0+\int_0^{\bar{r}_z}\mathrm d\bar{r}\left(\mathcal{A}-\mathcal{B}_\alpha n^\alpha-\mathcal{C}_{\alpha\beta}n^\alpha n^\beta\right)\,. \label{exdeltar}
\end{align} 
The quantity $\delta\hat\Gamma^0\equiv\hat\Gamma^0{_{\mu\nu}}\hat{\bar k}^\mu\hat{\bar k}^\nu$ can be calculated using the expressions for the Christoffel symbols given in Appendix~\ref{Appendix:JacobiMap}, which yields
\begin{align}
\delta\hat\Gamma^0=&\mathcal{A}'-2n^\alpha\mathcal{A}_{,\alpha}+\left(\mathcal{B}_{(\alpha,\beta)}+\mathcal{C}'_{\alpha\beta}\right)n^\alpha n^\beta\,, \nonumber \\
=&2\frac{\mathrm d}{\mathrm d\lambda}\left(\alpha_\chi+H\chi\right)-\left(\alpha_\chi-\varphi_\chi\right)'+\frac{\partial}{\partial\bar{r}}\left(\Psi_{\alpha}n^\alpha+C_{\alpha\beta}n^\alpha n^\beta\right) \nonumber \\
&+\frac{\mathrm d}{\mathrm d\lambda}C_{\alpha\beta}n^\alpha n^\beta+\frac{\mathrm d^2}{\mathrm d\lambda^2}\left(\frac{\chi}{a}\right)\,.
\end{align}
Furthermore, note that 
\begin{align}
\mathcal{A}-\mathcal{B}_\alpha n^\alpha-\mathcal{C}_{\alpha\beta}n^\alpha n^\beta=(\alpha_\chi-\varphi_\chi)-\Psi_\alpha n^\alpha-C_{\alpha\beta}n^\alpha n^\beta+\frac{\mathrm d}{\mathrm d\lambda}\left(\frac{\chi}{a}+\mathcal{G}_\alpha n^\alpha\right)\,,
\end{align}
which, combined with the expression for $\delta\hat\Gamma^0$, enables us to rewrite equation \eqref{exdeltar} as
\begin{align}
\delta r_\chi=&-n_\alpha\mathcal{G}^\alpha_s-\Delta\lambda_s+n_\alpha\left(\delta x^\alpha+\mathcal{G}^\alpha\right)_o+\bar{r}_z\left(\delta\nu+2H\chi+2\alpha_\chi+C_{\alpha\beta}n^\alpha n^\beta+\frac{\mathrm d}{\mathrm d\lambda}\left(\frac{\chi}{a}\right)\right)_o \nonumber \\
&-\int_0^{\bar{r}_z}\mathrm d\bar{r}\left(\alpha_\chi+\varphi_\chi+\Psi_\alpha n^\alpha+2C_{\alpha\beta}n^\beta n^\alpha\right)-2\int_0^{\bar{r}_z}\mathrm d\bar{r}\,H\chi \nonumber \\
&-\int_0^{\bar{r}_z}\mathrm d\bar{r}\,(\bar{r}_z-\bar{r})\left[(\alpha_\chi-\varphi_\chi)'-\frac{\partial}{\partial\bar{r}}\left(\Psi_{\alpha}n^\alpha+C_{\alpha\beta}n^\alpha n^\beta\right)\right]\,. \label{Deltar}
\end{align}
We can now use this equation to substitute $\Delta\lambda_s$ in the expression \eqref{kappaJMresult} for $\check\kappa$, which yields
\begin{align}
\check\kappa=&-\frac{\delta r_\chi}{\bar{r}_z}+\frac{1}{\bar{r}_z}n_\alpha\left(\delta x^\alpha+\mathcal{G}^\alpha\right)_o-\delta z_\chi+\left(\frac{3}{2}n^\alpha n^\beta C_{\alpha\beta}-n_\alpha V^\alpha\right)_o-\left(\varphi_\chi-\frac{1}{2}n^\alpha n^\beta C_{\alpha\beta}\right)_s \nonumber \\
&-\int_0^{\bar{r}_z}\mathrm d\bar{r}\,\left(\frac{\bar{r}_z-\bar{r}}{2\bar{r}_z\bar{r}}\right)\left(\widehat\nabla^2(\alpha_\chi-\varphi_\chi)-n^\alpha\widehat\nabla^2\Psi_\alpha-n^\alpha n^\beta\widehat\nabla^2 C_{\alpha\beta}\right) \nonumber \\
&+\int_0^{\bar{r}_z}\mathrm d\bar{r}\,\left(\frac{\bar{r}_z-\bar{r}}{2\bar{r}_z}\right)\frac{\partial}{\partial\bar{r}}\left(\Psi_\alpha n^\alpha+2C_{\alpha\beta}n^\alpha n^\beta\right)+\frac{1}{2\bar{r}_z}\int_0^{\bar{r}_z}\mathrm d\bar{r}\,\left(\widehat\nabla^\alpha\Psi_\alpha+n_\beta\widehat\nabla^\alpha C_{\alpha\beta}\right) \nonumber \\
&-\frac{1}{\bar{r}_z}\int_0^{\bar{r}_z}\mathrm d\bar{r}\left(\Psi_\alpha n^\alpha+2C_{\alpha\beta}n^\beta n^\alpha\right)\,, \label{newhatkappa}
\end{align}
where $\delta z_\chi=\delta z+H_s\chi_s$ and $\delta r_\chi=\delta r+n_\alpha\mathcal{G}^\alpha_s$ are gauge-invariant quantities as discussed in Section~\ref{Subsec:Shear}. Furthermore, we used the relation
\begin{align}
\widehat{\Delta\nu}_o=\left(\delta\nu+H\chi+\frac{\mathrm d}{\mathrm d\lambda}\left(\frac{\chi}{a}\right)+\alpha_\chi+n^\alpha V_\alpha\right)_o\,,
\end{align}
which is a rewritten form of equation \eqref{deltanu} for $\delta\nu_o$. Finally, we can use the relations
\begin{align}
&\widehat\nabla^2\left(n^\alpha\Psi_\alpha\right)=-2n^\alpha\Psi_\alpha+n^\alpha\widehat\nabla^2\Psi_\alpha+2\widehat\nabla^\alpha\Psi_\alpha\,, \nonumber \\
&\widehat\nabla^2\left(n^\alpha n^\beta C_{\alpha\beta}\right)=-2n^\alpha n^\beta C_{\alpha\beta}+n^\alpha n^\beta\widehat\nabla^2C_{\alpha\beta}+4\widehat\nabla^\beta\left(C_{\alpha\beta} n^\alpha\right)\,, \nonumber \\
&\frac{\partial}{\partial\bar{r}}\left(\Psi_\alpha n^\alpha+2C_{\alpha\beta}n^\alpha n^\beta\right)=-\frac{1}{\bar{r}}\widehat\nabla^\alpha\left(\Psi_\alpha+2C_{\alpha\beta}n^\beta\right)-\frac{2}{\bar{r}}C_{\alpha\beta}n^\alpha n^\beta\,,
\end{align}
to rewrite the expression \eqref{newhatkappa} for $\check\kappa$ into the expression~\eqref{checkkappa}.

\end{document}